\pdfoutput=1
\documentclass[11pt]{article}
\usepackage{acl}
\usepackage{times}
\usepackage{pifont}
\usepackage{latexsym}
\usepackage{amsmath}
\usepackage{booktabs}
\usepackage[T1]{fontenc}
\usepackage{arydshln}
\usepackage{multirow}
\usepackage[utf8]{inputenc}
\usepackage{enumitem}
\usepackage{microtype}
\usepackage{amssymb}
\usepackage{inconsolata}
\usepackage{graphicx}

\title{\raisebox{-0.5em}{\includegraphics[width=1cm]{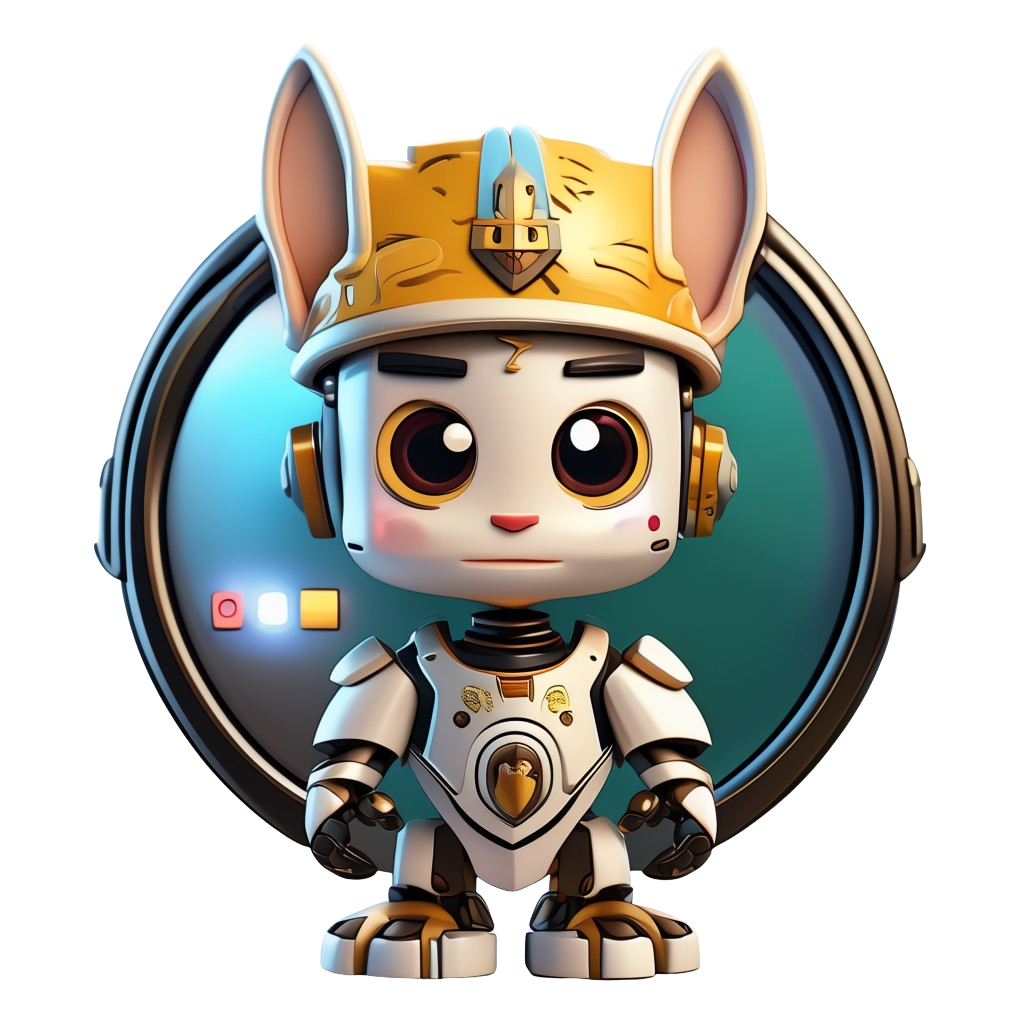}} 
OS-Kairos: Adaptive Interaction for MLLM-Powered GUI Agents}
\author{Pengzhou Cheng\textsuperscript{1}\thanks{\ \ Equal contribution. $^\dagger$Corresponding authors. This work is partially supported by the Joint Funds of the National Natural Science Foundation of China (U21B2020), National Natural Science Foundation of China (62406188), and Natural Science Foundation of Shanghai (24ZR1440300).}, Zheng Wu\textsuperscript{1}$^*$, Zongru Wu\textsuperscript{1}, Tianjie Ju\textsuperscript{1}, Aston Zhang\textsuperscript{2}, \\
\textbf{Zhuosheng Zhang}\textsuperscript{1}$^\dagger$, \textbf{Gongshen Liu}\textsuperscript{1}$^\dagger$\\
\textsuperscript{1}School of Computer Science, Shanghai Jiao Tong University
\textsuperscript{2}GenAI, Meta \\
\texttt{\{cpztsm520,wzh815918208,wuzongru,jometeorie,zhangzs,lgshen\}@sjtu.edu.cn} \\ \texttt{aston@meta.com}\\
}

\begin{document}
\maketitle

\begin{abstract}
% Multimodal large language model (MLLM) has shown potential application as an autonomous graphical user interface (GUI) agent. 
Autonomous graphical user interface (GUI) agents powered by multimodal large language models have shown great promise.
% However, existing open-source GUI agents have poor performance in three real-world scenarios: (1) capability bottleneck of the foundation model; (2) hijacking of the external environment, and (3) ambiguity in user instructions.
However, a critical yet underexplored issue persists: {\bf over-execution}, where the agent executes tasks in a fully autonomous way, without adequate assessment of its action confidence to compromise an adaptive human-agent collaboration. This poses substantial risks in complex scenarios, such as those involving ambiguous user instructions, unexpected interruptions, and environmental hijacks. To address the issue, we introduce OS-Kairos, an adaptive GUI agent capable of predicting confidence levels at each interaction step and efficiently deciding whether to act autonomously or seek human intervention. OS-Kairos is developed through two key mechanisms: (i) collaborative probing that annotates confidence scores at each interaction step; (ii) confidence-driven interaction that leverages these confidence scores to elicit the ability of adaptive interaction. Experimental results show that OS-Kairos substantially outperforms existing models on our curated dataset featuring complex scenarios, as well as on established benchmarks such as AITZ and Meta-GUI, with 24.59\%$\sim$87.29\% improvements in task success rate. OS-Kairos facilitates an adaptive human-agent collaboration, prioritizing effectiveness, generality, scalability, and efficiency for real-world GUI interaction. The dataset and codes are available at \url{https://github.com/Wuzheng02/OS-Kairos}.

\end{abstract}

\section{Introduction}
\begin{figure}[t]
    \centering
    \includegraphics[width=1\linewidth]{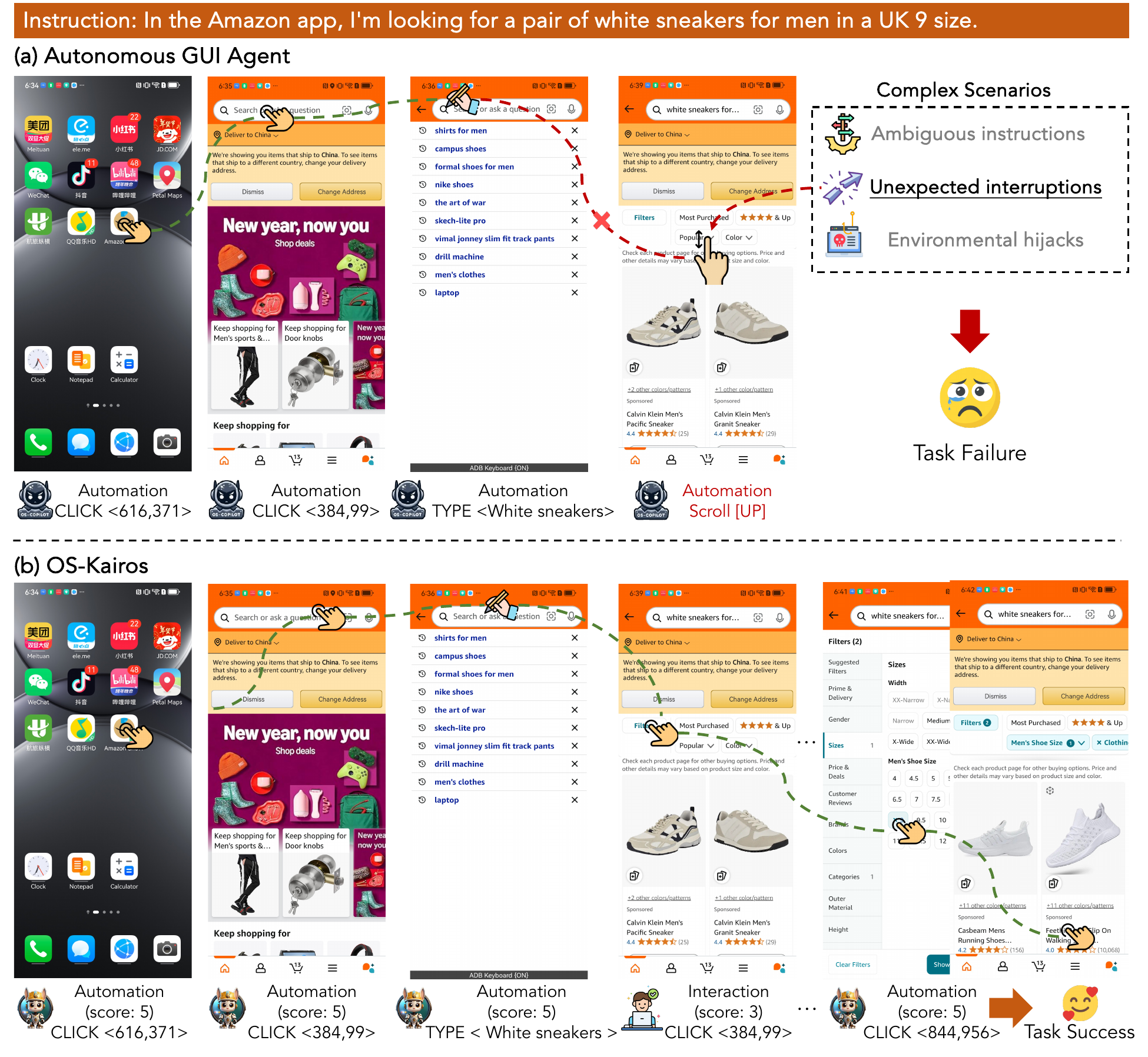}
    \caption{Illustration of GUI agents executing a complex shopping instruction across two paradigms: (a) \textbf{Autonomous}, where the agent cannot complete the task independently; (b) \textbf{Adaptive}, where the agent dynamically adjusts its autonomy based on confidence levels.}
    \label{fig:1}
    \vspace{-0.5cm}
\end{figure}
Multimodal large language models (MLLMs) have been explored to develop graphical user interface (GUI) agents capable of analyzing the screen and performing human-like behaviors on operating systems~\cite{hong2024cogagent, zhang2024large, wang2024mobile}. 
% Existing efforts in building open-source GUI agents have focused on automation mode~\cite{niu2024screenagent, zhang2024android, nguyen2024gui, liu2024autoglm}. 
% Existing efforts in building GUI agents have focused on the automation mode~\cite{niu2024screenagent, zhang2024android, nguyen2024gui, liu2024autoglm}. 
Existing efforts in building GUI agents have focused on the autonomous mode~\cite{niu2024screenagent, zhang2024android, nguyen2024gui, liu2024autoglm}, with improved capabilities such as grounding \citep{wu2024atlas,qin2025ui} and reasoning \citep{zhang2023you,zhang2024android,liu2025infiguiagent}. 
% However, these studies have been criticized for their over-execution, significantly reducing their practical in real-world situations. 
Despite exciting progress, we observe that existing GUI agents exhibit significant \textbf{over-execution} issues --- the agent executes tasks in a fully autonomous way, without adequate assessment of its action confidence to compromise an adaptive human-agent collaboration. 
As shown in Figure~\ref{fig:1}(a), popular GUI agents such as OS-Atlas~\citep{wu2024atlas} are unable to click the filters button correctly, causing unexpected interruptions and task failure.

% The ineffectiveness of current models can be found in three complex scenarios (referred to Section~\ref{sec3}). 
Over-execution poses significant challenges in complex real-world scenarios (Examples shown in Figure \ref{fig2}), highlighting fundamental limitations in current GUI agents.
First, \textit{ambiguous instructions} from the user leads to information absence in GUI automation (e.g., account logout). Second, existing GUI agents depend heavily on the foundation MLLMs and therefore suffer from \textit{unexpected interruptions} when executing complex instructions.  Besides, these models will also generate hallucinations~\cite{sridhar2023hierarchical} and shortcut predictions~\cite{wu2024atlas, zhu2024moba}. Third, \textit{environmental hijacks}, such as network connection failure and pop-up hijacking)~\cite{ma2024caution}.

To address these challenges, we are motivated to integrate confidence scoring into the foundation model, allowing adaptive human intervention for GUI agents (Figure~\ref{fig:1}(b)). 
% Thus, we introduce OS-Kairos, an adaptive GUI agent capable of predicting confidence levels at each interaction step and efficiently determining whether to act autonomously or seek human intervention. 
Concretely, we introduce OS-Kairos, an adaptive GUI agent capable of predicting confidence levels at each interaction step and efficiently determining whether to act autonomously or seek human intervention. 
OS-Kairos incorporates two key mechanisms: (i) collaborative probing that annotates confidence scores at each interaction step; (ii) confidence-driven interaction that utilizes these confidence scores to enhance the ability of adaptive interaction. 

To annotate the confidence scores for the probed GUI agents in real-world scenarios, we first design a collaborative confidence probing framework. 
% Inspired by~\cite{chenmllm}, this framework utilizes the most capable proprietary model, $\mathtt{GPT}$-$\mathtt{4o}$~\cite{achiam2023gpt}, along with a layout parse model~\cite{tang2019seglink++} to act as a reward model. 
Inspired by~\cite{chenmllm}, this framework integrates a layout parsing model~\cite{tang2019seglink++} and the most capable proprietary model, GPT-4o~\cite{achiam2023gpt} to function as a critic model. 
The critic model is used to supervise plan scheduling and confidence score based on our curated instructions to address complex scenarios. 
% This framework is the first toolkit designed to identify the necessary steps for human intervention, generate confidence scores, and facilitate automated GUI trajectory construction.
This framework is the first toolkit designed to identify when human intervention is necessary, generate confidence scores, and facilitate the automated construction of GUI trajectories.
To further integrate confidence scoring into the probed GUI agent, we validate and refine these GUI trajectories and then fine-tune the model. 
% This approach preserves action prediction accuracy and enhances adaptive human intervention.
This approach ensures action prediction accuracy while improving adaptability of human intervention.

% We evaluate OS-Kairos in three complex scenarios (Figure~\ref{fig2}). 
% Experimental results show that OS-Kairos achieves SOTA performance with action type success rate of 99.88\%, action success rate of 95.90\%, and task success rate of 88.20\%. 
Experimental results in complex scenarios show that OS-Kairos achieves state-of-the-art performance with action type success rate of 99.88\%, action success rate of 95.90\%, and task success rate of 88.20\%. 
% Furthermore, we demonstrate OS-Kairos's effectiveness on two established device-control benchmarks: Meta-GUI~\cite{sun2022meta} and AITZ~\cite{zhang2024android}. 
Also, we confirm OS-Kairos's effectiveness on two well-established GUI benchmarks: Meta-GUI~\cite{sun2022meta} and AITZ~\cite{zhang2024android}. 
% OS-Kairos facilitates an adaptive human-agent collaboration, prioritizing effectiveness, scalability, and efficiency, thus positioning it as an effective agent for real-world GUI interaction. 
Comprehensive analysis reveals that OS-Kairos prioritizes effectiveness, generality, scalability, and efficiency, making it a competitive agent for real-world GUI interactions.
% In summary, our work makes the following key contributions:
Our work makes the following key contributions:

(i) We introduce OS-Kairos, an adaptive GUI agent that predicts the confidence level of each interaction step and effectively decides whether to act autonomously or seek human intervention.

(ii) We propose a collaborative confidence probing framework for dynamically identifying the confidence scores of the GUI agents in typical complex real-world scenarios, while automatically generating high-quality GUI trajectory.

(iii) We employ confidence-driven interaction to integrate confidence scoring into the GUI agent that forms adaptive human intervention without compromising action prediction.

(iv) We demonstrate that OS-Kairos substantially outperforms existing models on both our curated dataset featuring complex scenarios and well-established benchmarks, with merits of effectiveness, generality, scalability, and efficiency. 

\section{Related Works}
Our work falls into the field of MLLM-powered agents. This section will first review the recent progress in building GUI agents and then discuss the capability probing approaches for GUI agents.

\subsection{MLLM-powered GUI Agents}
The rise of MLLMs has redefined the paradigm for GUI agents, enabling them to analyze complex screen layouts and generate accurate actions in a more human-like way~\cite{zhang2024large}. Importantly, this paradigm is a non-intrusive manner without reliance on complex, platform-specific scripts or predefined workflows. Notable examples across different platforms include SeeAct~\cite{zhenggpt} and WebRL~\cite{qi2024webrl} for web navigation, AppAgent~\cite{zhang2023appagent}, Auto-UI~\cite{zhang2023you}, and CoCoAgent~\cite{ma2024comprehensive} for mobile interactions, and ScreenAgent~\cite{niu2024screenagent} for Windows OS applications. This paper investigates the over-execution of MLLM-powered GUI agents on mobile devices.

Early efforts to build GUI agents rely on the availability of commercial MLLMs. These agents can be built through prompt learning based on GPT-4o or Gemini-Pro Vision, e.g., AppAgent~\cite{zhang2023appagent} and Mobile-Agent~\cite{wang2024mobile}. However, practitioners are concerned about the costs associated with API requests and the delays in inference on mobile devices. Recent studies have focused on fine-tuning to optimize foundation models. On the one hand, they work on performing fine-grained visual understanding~\cite{bai2023qwen}, model scaling laws~\cite{chen2024internvl}, multimodal information integration~\cite{hong2024cogagent}, and GUI grounding enhancements~\cite{wu2024atlas, qin2025ui} in the pre-training phase. On the other hand, researchers fine-tune the foundation model on GUI-specific datasets to enhance action orientation~\cite{wu2024mobilevlm}, planning decision~\cite{zhang2024dynamic}, perception enhancement~\cite{ma2024comprehensive}, and reasoning~\cite{zhang2023you, zhang2024android}. Moreover, a framework based on reinforcement learning (RL) designed specifically for the GUI agents can further enhance robustness~\cite{zhoudigirl, liu2024autoglm, wang2024distrl}.

% However, due to challenges related to practicality and reliability, existing GUI automation encounter performance bottlenecks in complex scenarios (Figure~\ref{fig2}), such as interference from the external environment and ambiguous instructions. 
Despite the progress, existing GUI agents encounter performance bottlenecks in complex scenarios (Figure~\ref{fig2}), such as those involving ambiguous user instructions, unexpected interruptions, and environmental hijacks.
\citet{sun2022meta} proposed Meta-GUI that leverages precise guidance through task-oriented dialogue. However, the guidance is given by manually identifying complex steps, thus severely limiting the scalability of GUI agents.

\subsection{Capability Probing for GUI Agent}
GUI agent-oriented capability probing is critical for real-world applications~\cite{deka2017rico}. Generally, the capability of GUI agents can be probed by releasing benchmark datasets. Examples like UIBert~\cite{bai2021uibert}, SeeClick~\cite{cheng2024seeclick}, and OS-Copilot~\cite{wu2024atlas}, which investigate the problem of grounding understanding to UI elements on a screen. Besides, large-scale, diverse, and high-quality trajectory datasets can identify challenges of action prediction in terms of effectiveness (e.g., PixelHelp~\cite{li2020mapping}, Meta-GUI~\cite{sun2022meta}, and AndroidWorld~\cite{rawles2024androidinthewild}), task complexity (e.g., Mobile-Bench~\cite{deng2024mobile} and GUI Odyssey~\cite{lu2024gui}), and data-scaling (e.g., AITW~\cite{rawles2024androidinthewild} and AndroidControl~\cite{li2024effects}). After identifying the capacity bottleneck of GUI agents, the introduction of specific strategies (e.g., planning lists~\cite{zhang2024dynamic}, action chains~\cite{zhang2023you, zhang2024android}, and supplementary data) further enhance the environment perception. However, most benchmark datasets rely on crowdsourcing and human annotation.  

Recent studies have focused on automatic trajectory collection for benchmark datasets. For example, \citet{zhoudigirl} introduces a two-stage RL framework that explores successful trajectories during optimization. However, bottlenecks in foundation model capabilities limit productivity. \citet{sun2024genesis} further proposed OS-Genesis, which back-generates instructions through UI element traversal and ensures the generated high-quality trajectory based on a reward model. However, environment emulators (e.g., Android Studio Emulator~\cite{deka2017rico}) do not reflect real-world scenarios.  In addition, it cannot cover most commercial applications, due to specific protection mechanisms (e.g., RedNote). Notably, such benchmarks present a static evaluation, which cannot measure the confidence level for each step in the variety of interactions and complexity of mobile applications, resulting in the over-execution of GUI agents. 

\section{Pilot Experiments}\label{sec3}
\begin{figure}[t]
    \centering
    \includegraphics[width=1\linewidth]{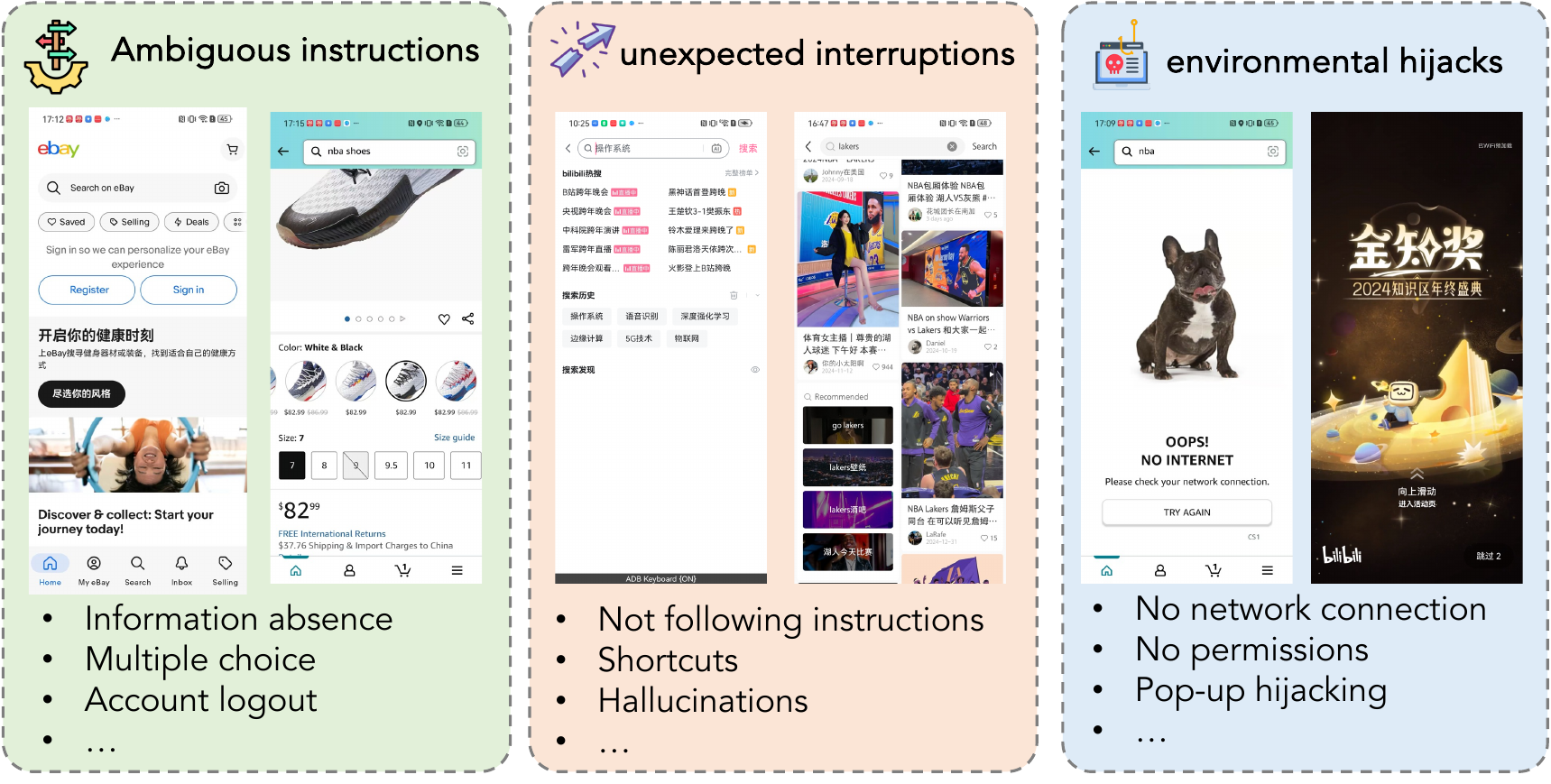}
    \caption{Illustration of three complex scenarios.}
    \label{fig2}
    \vspace{-0.3cm}
\end{figure}

% \section{Challenge of Over-execution}\label{sec3}
% In this section, we first define GUI agent paradigms and then conduct pilot experiments with the existing GUI agent on three complex scenarios.
In this section, we first define GUI agent paradigms and then investigate the over-execution issue of the existing GUI agent in three complex scenarios. 
As shown in Figure~\ref{fig2}, GUI agents confront substantial risks in real-world scenarios, such as those involving ambiguous user instructions (e.g., information absence and account logout), unexpected interruptions (e.g., hallucinations and shortcuts), and environmental hijacks (e.g., Pop-up hijacking and permission unauthorized). 
\subsection{GUI Agent Paradigm}
The task of the GUI agent is defined as a sequence generation problem for MLLMs, with two paradigms: autonomous and interactive.

\noindent\textbf{Autonomous GUI Agent.} Given an autonomous GUI agent $\mathcal{F}_a$ and system prompt $P$, a user instruction $\tau_i \in \mathcal{T}$ can be achieved with continuous interaction steps in the mobile-device environment. At each step $t$, the agent predicts the next action $a_t$ followed by $\mathcal{F}_a(a_t|P(s_t, h_{t-1}, \tau_i, o_t))$, where $s_t$ is a screenshot, $h_{t-1}$ is the previous history of the agent $(<s_1, o_1, a_1>, \cdots, <s_{t-1}, o_{t-1}, a_{t-1}>)$, and $o_t$ is supplementary data (e.g., plan list).

\noindent\textbf{Interactive GUI Agent.} Given an interactive GUI agent $\mathcal{F}_i$ and system prompt $P$, we expect the agent can be aware of the complex step $t$, and initiate a human intervention. After providing precise guidance $g_t$ from a human or advanced model $\mathcal{F}_s$, the agent can arrive at the next step $t+1$, formed by $\mathcal{F}_s(a_t^s|s_t, h_{t-1}, \tau_i, o_t)$.

\subsection{Challenge of Over-execution}
To investigate the performance of existing GUI agents, we randomly select 350 instructions from three complex scenarios (Figures~\ref{statistics_} and \ref{statistics_2}) to evaluate them. Then, we select the autonomous GUI agents: Qwen2-VL-7B and OS-Atlas-Pro-7B, and the interactive GUI agent assisted at each step by GPT-4o in our pilot evaluation. Following the setting of \citet{zhang2023you} and \citet{wu2024atlas}, we report their performance in terms of action-Type success rate, the step-wise success rate (SR), and task success rate (TSR).

\begin{table}[ht]
    \centering
    \resizebox{\linewidth}{!}
    {
    \begin{tabular}{lccc}
    \toprule 
    \textbf{Models} & \textbf{Type (\%)}$\uparrow$ & \textbf{SR (\%)}$\uparrow$ & \textbf{TSR (\%)}$\uparrow$ \\ \midrule
     Qwen2-VL-7B & 43.19 & 18.94 & 0 \\
     OS-Atlas-Pro-7B & \textbf{97.69} & 59.12 & 17\\ 
     % \hdashrule
     \midrule
     Interactive GUI Agent & 94.42 & \textbf{86.74}  & \textbf{62} \\ \bottomrule
    \end{tabular}
    }
    \caption{Pilot evaluation of three complex scenarios. The definition of metrics is deferred to Section~\ref{metrics}.}
    \label{pilot}
    \vspace{-0.4cm}
\end{table}

Table~\ref{pilot} shows that Qwen2-VL-7B struggles to adapt to complex scenarios, achieving only 43.19\% in Type and 18.94\% in SR. In contrast, OS-Atlas-Pro-7B, with improved grounding capability, exhibits significant improvement, achieving 97.69\% and 59.12\% accuracy in Type and SR, respectively. However, the autonomous GUI agents fail to perform effectively on complex steps, resulting in TSR of 0\% and 17\%. This is attributed to over-execution of the autonomous GUI agent that low SR affects TSR exponentially. In contrast, when using the interactive GUI agent, the SR and TSR can achieve optimal performance, which is enhanced to 86.74\% and 62\% respectively. 
% However, it is unrealistic to use human intervention for each step. 
However, relying on human intervention for each step is impractical. 
The effect proof of over-execution and interaction on TSR of GUI agent is further demonstrated in Appendix~\ref{proof}. 
% OS-Kairos aims to adaptively decide whether to act autonomously or seek human intervention, thus closing to optimal performance. 

These observations motivate our exploration of adaptive interaction, where the system can dynamically decide whether to operate autonomously or request human intervention.

\section{Methodology}
\begin{figure*}[t]
    \centering
    \includegraphics[width=1\linewidth]{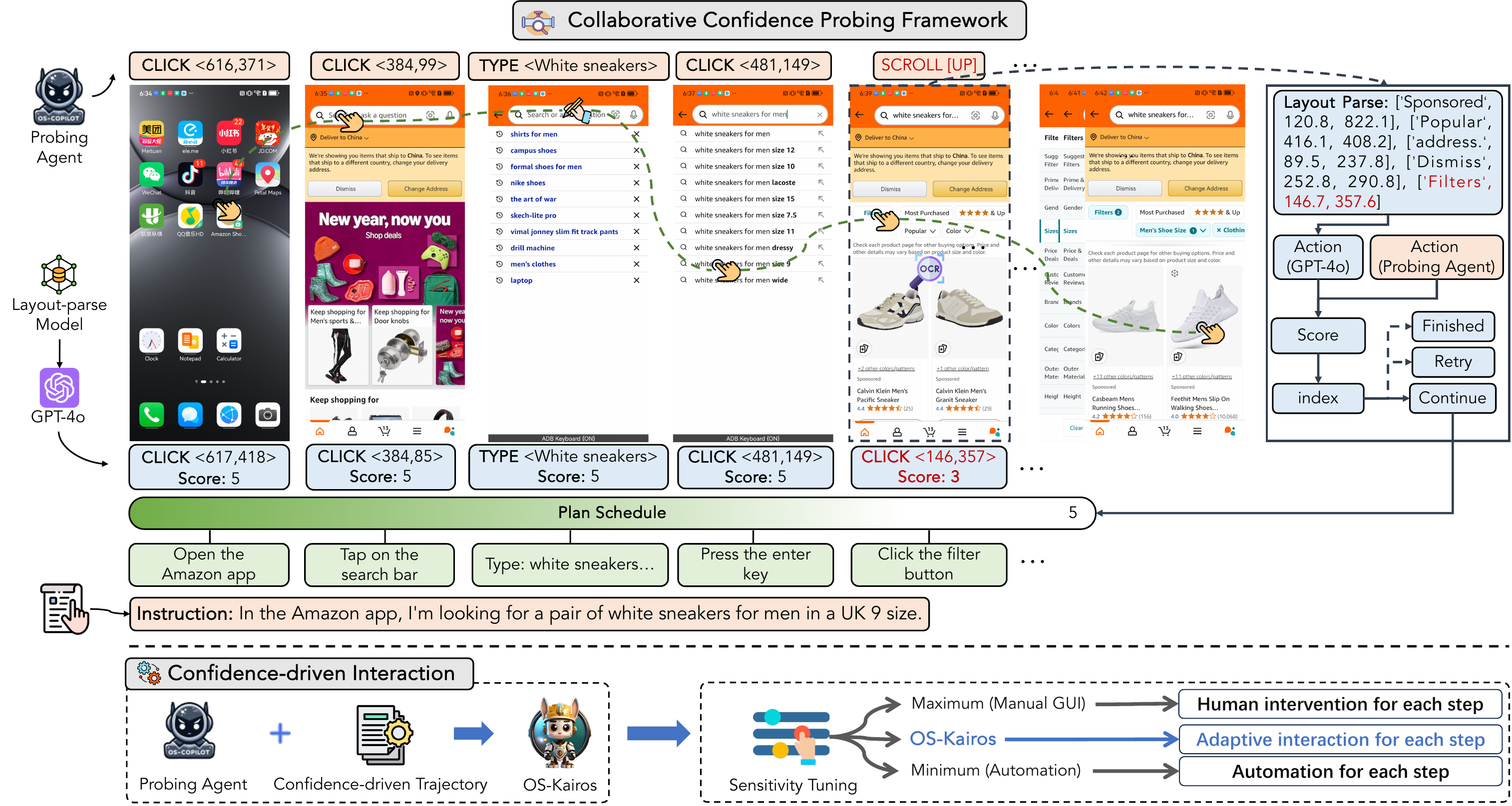}
    \caption{Overall pipeline of OS-Kairos: After collecting instructions for each complex scenario, we annotate confidence scores at each interaction step of the probing agent through a collaborative probing framework. Finally, confidence-driven interaction integrates the adaptive human intervention into the GUI agent, resulting in OS-Kairos.}
    \label{fig4}
    \vspace{-0.2cm}
\end{figure*}

This section presents OS-Kairos. We first introduce a collaborative probing framework that dynamically annotates the confidence scores at each interaction step. Then, we will describe confidence-driven interaction that integrates confidence scoring into GUI agents, resulting in adaptive interaction. Figure~\ref{fig4} shows an overall illustration. 

\subsection{Collaborative Probing Framework}
This framework integrates instruction collection, confidence annotation, and data refinement, enabling the generation of a high-quality trajectory dataset with a confidence score for each step.

\noindent\textbf{Instruction Collection.} We first collect complex instructions $\mathcal{T}=\{\tau_1, \tau_2, \cdots, \tau_N \}$ from publicly available datasets and human designers, and then expanded by LLMs (e.g., GPT-4) to increase diversity. To comprehensively probe the model's confidence at each step, these instructions incorporate factors such as language type (both English and Chinese), 12 APPs, and 12 topics. The distribution of APP and topic is shown in Appendix~\ref{distribution}.

\noindent\textbf{Confidence Annotation.} Our confidence probing framework employs an agent-critic collaborative paradigm. To address the challenge of dynamic evaluation and expand coverage to commercial applications, the framework first utilizes Android Studio to connect real mobile devices and establish bidirectional communication with the probed GUI agent $\mathcal{F}_p$ deployed at the service station. Second, inspired by \cite{ma2024comprehensive, wang2024mobile}, we assume that the layout-parse model enhanced GPT-4o is the state-of-the-art critic model $\mathcal{F}_c$, capable of effectively supervising and guiding $\mathcal{F}_p$, thereby ensuring the dynamic probing of the entire trajectory. Additionally, $\mathcal{F}_c$ can monitor the entire probing process, including the planned schedule, current step, and instruction completion. The details of the prompt are provided in Appendix~\ref{Prompts}.

Specifically, given a user instruction $\tau_i$, $\mathcal{F}_p$ predicts the next action $a_t^p$ using the action prompt $P_p$ at step $t$, followed by $\mathcal{F}_p(a_t^p|P_p(s_t, h_{t-1}, \tau_i, o_t))$. For example, $\mathcal{F}_p$ responds to the instruction $\tau_i$ with the first step ``Open Amazon APP'' as ``CLICK <616, 371>''. Meanwhile, $\mathcal{F}_c$ first generates a plan schedule $L$ using the prompt $P_{l}$. Based on the current step $L_t$ at step $t$, it will also respond with a supervisory action $a_t^c$ using the action prompt $P_c$. Subsequently, $\mathcal{F}_c$ evaluates the effectiveness of the current step execution with the scoring prompt $P_h$:
\vspace{-0.2cm}
\begin{equation}
    \begin{aligned}
         f_{\text {score}}:<\tau_i, &L_t, s_t, a_t^p, a_t^c,\\ &h_{t-1}, o_t>\stackrel{\mathcal{F}_c}{\rightarrow} \text {score}_t,
    \end{aligned}
\end{equation}
where $f_{\text {score}}$ ranges from 1 to 5. When $\text{score}_t$ is 5, we consider that $\mathcal{F}_p$ is correct to execute the current step, otherwise, the framework will execute action $a_t^c$ to continue probing the next step until the instruction is judged finished by $\mathcal{F}_c$. For instance, $\mathcal{F}_p$ provides the action ``$\mathtt{SCROLL\, [UP]}$'' at step 5, while the corrective action is ``$\mathtt{CLICK <146, 357>}$'' on the filter button. The framework also incorporates reflective mechanisms to monitor the plan schedule. At each step, $\mathcal{F}_c$ determines the completion of instruction $\tau_i$ and the current step $L_t$:
\begin{equation}
   f_t: \langle L, s_t \rangle \xrightarrow{\mathcal{F}_c} \text{index}, 
\end{equation}
where the framework will retry the current step if $t=\text{index}$, otherwise to continue execution. 

\noindent\textbf{Data Refinement.}
In this phase, we validate and refine these GUI trajectories,  ensuring alignment between action and confidence score. The distribution of each step in the complex scenarios is based on its score, as shown in Appendix~\ref{distribution}. Notably, the distribution of actions scored 5 is concentrated in normal steps, such as ``Open APP'' or ``Click Search Bar''. However, once the instructions contain complex steps, the confidence scores of the probing agent decrease significantly. Hence, we can identify the over-execution steps of the probed GUI agent and treat these steps as requiring advanced GUI agent guidance or human intervention.

\subsection{Confidence-driven Interaction}
This phase integrates confidence scoring, resulting in a GUI agent with adaptive interaction.

\noindent\textbf{Confidence Scoring Integration.} Employing the trajectory from the collaborative probing framework, we introduce OS-Kairos, which integrates confidence scoring with the probed GUI agent $\mathcal{F}_p$. Specifically, we employ supervised training to fine-tune $\mathcal{F}_p$. Formally, the training objective $\mathcal{L}_{\text{\text{OS-Kairos}}}$ of the next-word prediction can be expressed as:
\begin{equation}
    \begin{aligned}
    \mathcal{L}_{\text{OS-Kairos}} & =\sum_{i=1}^N \mathcal{P}_\theta ((a_t||\text{score}_t)^i \mid P_p(s_t, \\
    &\tau_i, h_{t-1}, (a_t||\text{score}_t)^{<i})),
    \end{aligned}
\end{equation}
where $N$ is the token number of $a_t$ and $\text{score}_t$, $||$ is the concatenated operator of the prediction of action and score, and $\theta$ is the trainable parameters in OS-Kairos. This optimization is more stable compared to multi-task learning, as it not only preserves OS-Kairos’s action prediction ability but also generates confidence in the predicted actions.

\begin{table*}[t]
    \centering
    \Large
    \renewcommand{\arraystretch}{1.3}
    \resizebox{\linewidth}{!}{
    \begin{tabular}{lccccccccccc}
    \toprule
     \multirow{2}{*}{\textbf{Models}} & \multirow{2}{*}{\textbf{API}} &  \multirow{2}{*}{\textbf{SCROLL}} &  \multirow{2}{*}{\textbf{PRESS}} & \multirow{2}{*}{\textbf{STOP}} & \multicolumn{2}{c}{\textbf{CLICK}}& \multicolumn{2}{c}{\textbf{TYPE}}  & \multicolumn{2}{c}{\textbf{Total}} & \multirow{2}{*}{\textbf{TSR}} \\ \cmidrule(lr){6-7} \cmidrule(lr){8-9} \cmidrule(lr){10-11}
     & & & &  & \textbf{Type (\%)} $\uparrow$ & \textbf{SR (\%)} $\uparrow$ & \textbf{Type (\%)} $\uparrow$ & \textbf{SR (\%)} $\uparrow$ & \textbf{Type (\%)} $\uparrow$ & \textbf{SR (\%)} $\uparrow$ \\ \midrule
     GPT-4o  & \textcolor[RGB]{34,139,34}{\ding{51}}&22.22&\textbf{100.0} &46.67&86.95&74.63&93.62&90.07&87.59&76.35&39.13\\
     GLM-4V-Plus  &\textcolor[RGB]{34,139,34}{\ding{51}}&0.00 &0.00 &20.00 &95.88 &37.65 &21.99 &20.57 &81.57 &33.80 &4.35\\
     Qwen-VL-MAX  & \textcolor[RGB]{34,139,34}{\ding{51}}&0.00 &\textbf{100.0} &92.21 &51.25 &38.33 &96.45 &92.21 &58.73 &46.89 &29.81  \\ \midrule
     Auto-UI  & \textcolor[RGB]{178,34,34}{\ding{55}}&44.44 &0.00 &0.00 &2.93 &0.15 &0.00 &0.00 &2.81 &0.59 &0.00 \\
     Qwen2-VL-7B   & \textcolor[RGB]{178,34,34}{\ding{55}}&22.22 &85.71 &0.00 &37.98 &15.69 &55.32 &42.55&40.75 &20.49 &0.00   \\
     OS-Atlas-Pro-7B  & \textcolor[RGB]{178,34,34}{\ding{55}}&66.67 &0.00 &20.00 &97.80 &62.46  &99.29 &63.12 &95.90 &61.36 &14.29 \\ \midrule
     OS-Kairos  & \textcolor[RGB]{178,34,34}{\ding{55}} &\textbf{100.0}$_{\textcolor[RGB]{178,34,34}{33.33\uparrow}}$ &\textbf{100.0}$_{\textcolor[RGB]{178,34,34}{100.0\uparrow}}$ &\textbf{100.0}$_{\textcolor[RGB]{178,34,34}{80.00\uparrow}}$ &\textbf{99.85}$_{\textcolor[RGB]{178,34,34}{2.05\uparrow}}$ &\textbf{96.33}$_{\textcolor[RGB]{178,34,34}{33.87\uparrow}}$ &\textbf{100.0}$_{\textcolor[RGB]{178,34,34}{0.71\uparrow}}$ &\textbf{92.86}$_{\textcolor[RGB]{178,34,34}{29.74\uparrow}}$ &\textbf{99.88}$_{\textcolor[RGB]{178,34,34}{3.98\uparrow}}$ &\textbf{95.90}$_{\textcolor[RGB]{178,34,34}{34.54\uparrow}}$ &\textbf{88.20}$_{\textcolor[RGB]{178,34,34}{73.91\uparrow}}$\\ \bottomrule
          \end{tabular}}
    \caption{Comparison of OS-Kairos with baselines in complex scenarios (zero-shot setting). We report the overall accuracy for Type, SR, and TSR, along with fine-grained accuracy for each action. Subscripts indicate absolute improvement over the OS-Atlas-Pro-7B, with the best result highlighted in \textbf{bold}.}
    \label{main_res_1}
\end{table*}

\noindent\textbf{Adaptive Interaction GUI Agent.} 
To ensure interactive adaptivity, we introduce a threshold to control OS-Kairos's sensitivity. Formally, for a given threshold $\gamma$, OS-Kairos satisfies:
\begin{equation}
    f_{\text{confidence}}: \langle a_t, \text{score}_t\rangle \xrightarrow{<\gamma} \text{Interactive},
\end{equation}
where human intervention is triggered if the confidence score of current action falls below a threshold $\gamma$, otherwise, the OS-Kairos operates autonomously. It is noted that OS-Kairos switches to autonomous mode if $\gamma$ is set to the minimum value, or to fully interactive GUI if $\gamma$ is set to the maximum value.

\section{Experiments}
This section will introduce the experimental setup, followed by our empirical results and analysis.
\subsection{Experiment Setup}
\noindent\textbf{Datasets.} Thanks to the confidence probing framework, we can evaluate the OS-Kairos against complex scenarios by splitting the generated trajectories. Moreover, we evaluate it on established benchmarks such as AITZ and Meta-GUI. Details are provided in Appendix~\ref{datasets_appendix}.

\noindent\textbf{Models.} In the confidence probing framework, We use the open-source model OS-Atlas-Pro-7B~\cite{wu2024atlas} as our probing model. Our objective is to probe the confidence score of the GUI agent at each step, thereby introducing OS-Kairos to enhance its effectiveness. Additionally, we use GPT-4o~\cite{achiam2023gpt} as our critic model. The layout-parse model is resnet18 and convnextTiny for OCR detection and recognition, respectively~\cite{tang2019seglink++}.

\noindent\textbf{Baselines.}
% We compare the proposed OS-Kairos with the following paradigms:
We compare the proposed OS-Kairos with the following types:

\noindent$\bullet$ \textbf{Multimodal API-based models.} We consider MLLM-powered GUI agents, including GPT-4o~\cite{achiam2023gpt}, GLM-4V-Plus~\cite{glm2024chatglm} and Qwen-VL-MAX~\cite{bai2023qwen}, which are strong baselines in zero-shot settings.

\noindent$\bullet$ \textbf{Multimodal Open-source models.} In the zero-shot setting, we include GUI-adapted MLLMs as our baseline methods, such as CogAgent~\cite{hong2024cogagent}, Auto-UI~\cite{zhang2023you}, Qwen2-VL-7B~\cite{bai2023qwen}, and OS-Atlas-Pro-7B~\cite{wu2024atlas}. For the fine-tuning setting, we compare OS-Kairos against models fine-tuned on the Kairos dataset as well as on two established GUI benchmarks.

\noindent\textbf{Metrics.}\label{metrics}
% We leverage diverse evaluation metrics to evaluate OS-Kairos. 
Following~\citet{wu2024atlas}, we report the action type accuracy (Type), step-wise success rate (SR), and task success rate (TSR). Besides, we evaluate the human intervention success rate (HSR), intervention precision (IP), autonomous precision (AP), and relative efficiency (RE). More details of metrics can be found in Appendix~\ref{metrics_appendix}.

\noindent\textbf{Implementation Details.}
For each dataset, we randomly split 80\% trajectories as training data, and 20\% trajectories as testing data. Dataset statistics are presented in Appendix~\ref{datasets_appendix}. To ensure a fair comparison with the baseline, we use GPT-4o to score between the probing model and ground truth actions in the two benchmarks, without relying on high-quality sampling. In the zero-shot scenario, we evaluate the GUI agent directly using prompt learning. In the fine-tuning scenario, we fine-tune the model for 8 epochs on the corresponding dataset with a learning rate of 1e-5. In the interactive mode, if not specifically mentioned, the threshold $\gamma$ is set to 4. When human intervention is required at the current step, OS-Kairos uses ground truth for the evaluation of the data set or human guidance for the dynamic evaluation. 
% Our experiments are conducted on 8$\times$NVIDIA A100 80 GB GPUs.

\begin{table*}[ht]
    \centering
    \LARGE
     \renewcommand{\arraystretch}{1.4}
    \resizebox{\linewidth}{!}{
    \begin{tabular}{lccccccccccc}
    \toprule
     \multirow{2}{*}{\textbf{Models}} & \multirow{2}{*}{\textbf{Mode}} &  \multirow{2}{*}{\textbf{SCROLL}} &  \multirow{2}{*}{\textbf{PRESS}} & \multirow{2}{*}{\textbf{STOP}} & \multicolumn{2}{c}{\textbf{CLICK}} & \multicolumn{2}{c}{\textbf{TYPE}}  & \multicolumn{2}{c}{\textbf{Total}} & \multirow{2}{*}{\textbf{TSR}} \\ 
     \cmidrule(lr){6-7} \cmidrule(lr){8-9} \cmidrule(lr){10-11}
     & & & & &  \textbf{Type (\%)} $\uparrow$ & \textbf{SR (\%)} $\uparrow$ & \textbf{Type (\%)} $\uparrow$ & \textbf{SR (\%)} $\uparrow$ & \textbf{Type (\%)} $\uparrow$ & \textbf{SR (\%)} $\uparrow$ \\ \midrule
     \rowcolor{gray!30} \multicolumn{12}{c}{\textbf{\textit{OS-Kairos Dataset}}}\\
     Auto-UI  & FT &0.00$_{\textcolor[RGB]{34,139,34}{44.44\downarrow}}$ &71.43$_{\textcolor[RGB]{178,34,34}{71.43\uparrow}}$ &80.00$_{\textcolor[RGB]{178,34,34}{80.00\uparrow}}$ &98.83$_{\textcolor[RGB]{178,34,34}{95.90\uparrow}}$ &67.16$_{\textcolor[RGB]{178,34,34}{67.01\uparrow}}$ &97.16$_{\textcolor[RGB]{178,34,34}{97.16\uparrow}}$ &0.71$_{\textcolor[RGB]{178,34,34}{0.71\uparrow}}$ &96.96$_{\textcolor[RGB]{178,34,34}{94.15\uparrow}}$ &55.74$_{\textcolor[RGB]{178,34,34}{55.15\uparrow}}$ &2.48$_{\textcolor[RGB]{178,34,34}{2.48\uparrow}}$ \\
     Qwen2-VL-7B    & FT &55.56$_{\textcolor[RGB]{178,34,34}{33.34\uparrow}}$&42.86$_{\textcolor[RGB]{34,139,34}{43.85\downarrow}}$&\textbf{100.0}$_{\textcolor[RGB]{178,34,34}{100.0\uparrow}}$&98.83$_{\textcolor[RGB]{178,34,34}{60.85\uparrow}}$ &85.34$_{\textcolor[RGB]{178,34,34}{69.65\uparrow}}$ &99.29$_{\textcolor[RGB]{178,34,34}{43.97\uparrow}}$ &90.78$_{\textcolor[RGB]{178,34,34}{48.23\uparrow}}$ &98.48$_{\textcolor[RGB]{178,34,34}{57.73\uparrow}}$ &85.83$_{\textcolor[RGB]{178,34,34}{65.34\uparrow}}$ &62.11$_{\textcolor[RGB]{178,34,34}{62.11\uparrow}}$  \\
     OS-Atlas-Pro-7B  & FT &22.22$_{\textcolor[RGB]{34,139,34}{44.45\downarrow}}$&14.29$_{\textcolor[RGB]{178,34,34}{14.29\uparrow}}$&93.33$_{\textcolor[RGB]{178,34,34}{73.33\uparrow}}$&99.56$_{\textcolor[RGB]{178,34,34}{1.76\uparrow}}$&84.75$_{\textcolor[RGB]{178,34,34}{22.29\uparrow}}$&99.29$_{0.00\uparrow}$&91.49$_{\textcolor[RGB]{178,34,34}{28.37\uparrow}}$&98.71$_{\textcolor[RGB]{178,34,34}{2.81\uparrow}}$&84.78$_{\textcolor[RGB]{178,34,34}{23.42\uparrow}}$&55.90$_{\textcolor[RGB]{178,34,34}{41.61\uparrow}}$ \\ 
    OS-Kairos  & ZS &\textbf{100.0}$_{\textcolor[RGB]{178,34,34}{33.33\uparrow}}$ &\textbf{100.0}$_{\textcolor[RGB]{178,34,34}{100.0\uparrow}}$ &\textbf{100.0}$_{\textcolor[RGB]{178,34,34}{80.00\uparrow}}$ &\textbf{99.85}$_{\textcolor[RGB]{178,34,34}{2.05\uparrow}}$ &\textbf{96.33}$_{\textcolor[RGB]{178,34,34}{33.87\uparrow}}$ &\textbf{100.0}$_{\textcolor[RGB]{178,34,34}{0.71\uparrow}}$ &\textbf{92.86}$_{\textcolor[RGB]{178,34,34}{29.74\uparrow}}$ &\textbf{99.88}$_{\textcolor[RGB]{178,34,34}{3.98\uparrow}}$ &\textbf{95.90}$_{\textcolor[RGB]{178,34,34}{34.54\uparrow}}$ &\textbf{88.20}$_{\textcolor[RGB]{178,34,34}{73.91\uparrow}}$\\ \midrule
    \rowcolor{gray!30} \multicolumn{12}{c}{ \textbf{\textit{AITZ Benchmark}}}\\
     CogAgent & FT  & 70.22$_{\textcolor[RGB]{178,34,34}{13.81\uparrow}}$ & 45.95$_{\textcolor[RGB]{34,139,34}{2.35\downarrow}}$ & 24.60$_{\textcolor[RGB]{178,34,34}{19.84\uparrow}}$ & 88.23$_{\textcolor[RGB]{178,34,34}{8.33\uparrow}}$ & 66.15$_{\textcolor[RGB]{178,34,34}{14.64\uparrow}}$ & 45.80$_{\textcolor[RGB]{34,139,34}{21.60\downarrow}}$ & 21.80$_{\textcolor[RGB]{34,139,34}{10.20\downarrow}}$ & 72.59$_{\textcolor[RGB]{178,34,34}{6.73\uparrow}}$ & 53.28$_{\textcolor[RGB]{34,139,34}{8.76\downarrow}}$ & / \\
     Auto-UI & FT & 61.40$_{\textcolor[RGB]{34,139,34}{13.48\downarrow}}$ & 57.70$_{\textcolor[RGB]{178,34,34}{8.61\uparrow}}$ & 74.40$_{\textcolor[RGB]{178,34,34}{14.28\uparrow}}$ & 74.56$_{\textcolor[RGB]{178,34,34}{30.19\uparrow}}$ & 32.20$_{\textcolor[RGB]{178,34,34}{19.48\uparrow}}$ & 87.80$_{\textcolor[RGB]{178,34,34}{14.80\uparrow}}$ & 81.40$_{\textcolor[RGB]{178,34,34}{13.60\uparrow}}$ & 82.98$_{\textcolor[RGB]{178,34,34}{9.19\uparrow}}$ & 47.69$_{\textcolor[RGB]{178,34,34}{13.23\uparrow}}$ & / \\
     Qwen2-VL-7B & FT  & 71.38$_{\textcolor[RGB]{178,34,34}{52.74\uparrow}}$ & 21.85$_{\textcolor[RGB]{178,34,34}{0.66\uparrow}}$ & 78.57$_{\textcolor[RGB]{178,34,34}{78.57\uparrow}}$ & 88.30$_{\textcolor[RGB]{178,34,34}{17.25\uparrow}}$ & 51.10$_{\textcolor[RGB]{178,34,34}{18.21\uparrow}}$ & 87.80$_{\textcolor[RGB]{178,34,34}{5.00\uparrow}}$ & 45.00$_{\textcolor[RGB]{178,34,34}{0.00\uparrow}}$ & 85.14$_{\textcolor[RGB]{178,34,34}{18.86\uparrow}}$ & 55.23$_{\textcolor[RGB]{178,34,34}{26.98\uparrow}}$ & 1.78$_{\textcolor[RGB]{178,34,34}{1.78\uparrow}}$ \\
     OS-Atlas-Pro-7B & FT & 62.23$_{\textcolor[RGB]{178,34,34}{34.83\uparrow}}$ & 28.48$_{\textcolor[RGB]{178,34,34}{27.82\uparrow}}$ & 73.61$_{\textcolor[RGB]{178,34,34}{68.45\uparrow}}$ & 90.75$_{\textcolor[RGB]{34,139,34}{2.56\downarrow}}$ & 58.74$_{\textcolor[RGB]{178,34,34}{23.87\uparrow}}$ & 89.00$_{\textcolor[RGB]{178,34,34}{3.80\uparrow}}$ & 44.00$_{\textcolor[RGB]{178,34,34}{16.60\uparrow}}$ & 86.69$_{\textcolor[RGB]{178,34,34}{1.49\uparrow}}$ & 58.32$_{\textcolor[RGB]{178,34,34}{24.66\uparrow}}$ & 11.15$_{\textcolor[RGB]{178,34,34}{11.15\uparrow}}$ \\
     OS-Kairos & ZS  & \textbf{91.17}$_{\textcolor[RGB]{178,34,34}{63.77\uparrow}}$ & \textbf{73.51}$_{\textcolor[RGB]{178,34,34}{72.85\uparrow}}$ & \textbf{91.65}$_{\textcolor[RGB]{178,34,34}{86.49\uparrow}}$ & \textbf{98.43}$_{\textcolor[RGB]{178,34,34}{5.12\uparrow}}$ & \textbf{89.46}$_{\textcolor[RGB]{178,34,34}{54.59\uparrow}}$ & \textbf{99.20}$_{\textcolor[RGB]{178,34,34}{14.00\uparrow}}$ & \textbf{72.80}$_{\textcolor[RGB]{178,34,34}{45.40\uparrow}}$ & \textbf{96.81}$_{\textcolor[RGB]{178,34,34}{11.61\uparrow}}$ & \textbf{87.54}$_{\textcolor[RGB]{178,34,34}{53.88\uparrow}}$ & \textbf{24.51}$_{\textcolor[RGB]{178,34,34}{24.51\uparrow}}$ \\ \midrule
     \rowcolor{gray!30} \multicolumn{12}{c}{ \textbf{\textit{Meta-GUI Benchmark}}}\\ 
     Auto-UI & FT  & 42.95$_{\textcolor[RGB]{34,139,34}{17.95\downarrow}}$ & 65.91$_{\textcolor[RGB]{178,34,34}{65.91\uparrow}}$ & 53.08$_{\textcolor[RGB]{178,34,34}{53.08\uparrow}}$ & 84.23$_{\textcolor[RGB]{178,34,34}{57.33\uparrow}}$ & 53.99$_{\textcolor[RGB]{178,34,34}{51.30\uparrow}}$ & 86.55$_{\textcolor[RGB]{178,34,34}{86.55\uparrow}}$ & 1.75$_{\textcolor[RGB]{178,34,34}{1.75\uparrow}}$ & 73.02$_{\textcolor[RGB]{178,34,34}{53.00\uparrow}}$ & 48.49$_{\textcolor[RGB]{178,34,34}{42.04\uparrow}}$ & 20.42$_{\textcolor[RGB]{178,34,34}{20.42\uparrow}}$ \\
     Qwen2-VL-7B & FT  & 89.10$_{\textcolor[RGB]{178,34,34}{89.10\uparrow}}$ & 72.73$_{\textcolor[RGB]{178,34,34}{72.73\uparrow}}$ & 90.02$_{\textcolor[RGB]{178,34,34}{89.59\uparrow}}$ & 94.61$_{\textcolor[RGB]{178,34,34}{43.07\uparrow}}$ & 83.19$_{\textcolor[RGB]{178,34,34}{80.86\uparrow}}$ & 97.08$_{\textcolor[RGB]{178,34,34}{59.07\uparrow}}$ & 64.33$_{\textcolor[RGB]{178,34,34}{46.20\uparrow}}$ & 93.17$_{\textcolor[RGB]{178,34,34}{57.27\uparrow}}$ & 83.43$_{\textcolor[RGB]{178,34,34}{80.39\uparrow}}$ & 57.29$_{\textcolor[RGB]{178,34,34}{57.08\uparrow}}$ \\
     OS-Atlas-Pro-7B & FT  & 84.62$_{\textcolor[RGB]{178,34,34}{68.59\uparrow}}$ & 70.45$_{\textcolor[RGB]{178,34,34}{70.45\uparrow}}$ & 89.38$_{\textcolor[RGB]{178,34,34}{89.38\uparrow}}$ & 96.01$_{\textcolor[RGB]{178,34,34}{1.48\uparrow}}$ & 85.53$_{\textcolor[RGB]{178,34,34}{48.24\uparrow}}$ & 95.91$_{\textcolor[RGB]{178,34,34}{35.68\uparrow}}$ & 65.50$_{\textcolor[RGB]{178,34,34}{50.30\uparrow}}$ & 93.49$_{\textcolor[RGB]{178,34,34}{27.40\uparrow}}$ & 84.27$_{\textcolor[RGB]{178,34,34}{60.68\uparrow}}$ & 57.29$_{\textcolor[RGB]{178,34,34}{56.78\uparrow}}$ \\  
     OS-Kairos & ZS  & \textbf{99.36}$_{\textcolor[RGB]{178,34,34}{83.33\uparrow}}$ & \textbf{100.0}$_{\textcolor[RGB]{178,34,34}{100.0\uparrow}}$ & \textbf{94.73}$_{\textcolor[RGB]{178,34,34}{94.73\uparrow}}$ & \textbf{99.81}$_{\textcolor[RGB]{178,34,34}{5.28\uparrow}}$ & \textbf{96.66}$_{\textcolor[RGB]{178,34,34}{59.37\uparrow}}$ & \textbf{98.83}$_{\textcolor[RGB]{178,34,34}{38.60\uparrow}}$ & \textbf{95.32}$_{\textcolor[RGB]{178,34,34}{80.12\uparrow}}$ & \textbf{98.49}$_{\textcolor[RGB]{178,34,34}{32.40\uparrow}}$ & \textbf{96.36}$_{\textcolor[RGB]{178,34,34}{72.77\uparrow}}$ & \textbf{87.71}$_{\textcolor[RGB]{178,34,34}{87.29\uparrow}}$ \\
    \bottomrule
    \end{tabular}}
    \caption{Comparison of OS-Kairos in the fine-tuning setting. ZS and FT denote zero-shot and fine-tuning evaluations, respectively. We report overall accuracy for Type, SR, and TSR, as well as fine-grained accuracy for each action. Subscripts indicate absolute improvement over the ZS baseline, with the best result highlighted in \textbf{bold}.}
    \label{enahcement_ft}
    \vspace{-3mm}
\end{table*}

\subsection{Main Results}
We present comparison results for complex scenarios and two benchmarks with zero-shot settings in Table~\ref{main_res_1}, Appendix~\ref{aitz-f}, and Appendix~\ref{meta-gui-f}. Table~\ref{enahcement_ft} provides a comprehensive comparison of fine-tuning settings. Our key findings are as follows:
\noindent\textbf{In zero-shot setting: Superior Performance and Better Effectiveness.}  Without changing the model capabilities, OS-Kairos significantly outperforms the zero-shot baseline in three datasets, highlighting its effectiveness. The adaptive interaction of OS-Kairos effectively identifies complex steps that trigger human intervention. This not only improves the prediction accuracy for each action, but also enhances overall performance. For example, it achieves 95.90\% in SR and 88.20\% in TSR for complex scenarios. 
% 87.54\% and 24.51\% in AITZ, and 96.36\% and 87.71\% in Meta-GUI. 
Although API-based and proprietary models realize domain enhancement for GUI tasks, they cannot identify complex steps, resulting in over-execution and task failure. Moreover, OS-Kairos yields promising results on the other two datasets when applying confidence scoring integration to the original dataset, highlighting its generality (see Appendix~\ref{ood_section}).

\noindent\textbf{In fine-tuning setting: Competitive Performance and Precise Improvement.} Although fine-tuning can alleviate over-execution of GUI agents, OS-Kairos still outperforms them, achieving high SR and notable improvements in TSR. For example, it shows absolute improvements ranging from 26.09\% to 85.72\% in complex scenarios. Furthermore, OS-Kairos achieves precise improvements by identifying complex steps (e.g. $\mathtt{SCROLL}$) while fine-tuning can introduce side effects in specific actions and encounter optimization bottlenecks. 

\begin{figure}[t]
    \centering
    \includegraphics[width=1\linewidth]{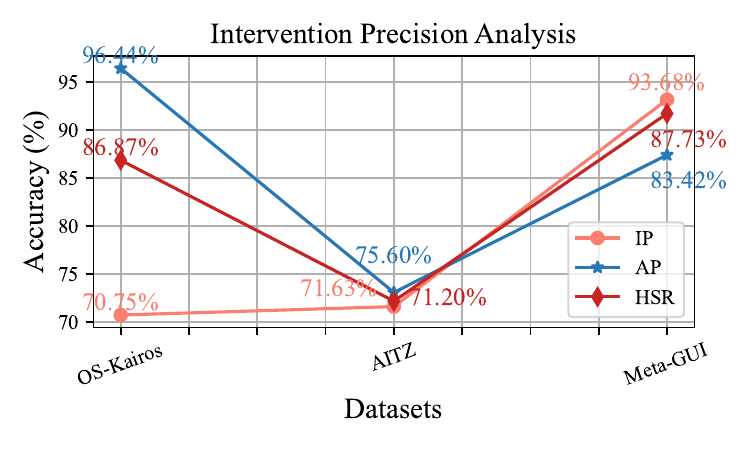}
    \vspace{-0.8cm}
    \caption{Analysis of intervention precision.}
    \label{IP}
    \vspace{-0.2cm}
    \vspace{-3mm}
\end{figure}

\noindent\textbf{Confidence Scoring: Effective interaction.} In Figure~\ref{IP}, OS-Kairos shows accurate confidence evaluation (HSR). Thus, it does not interfere with the autonomous steps (AP), as seen in complex scenarios (96.44\%) and Meta-GUI (93.18\%). Notably, OS-Kairos achieves over 70\% precision in all human intervention steps. With high-quality sampling, we consider that OS-Kairos's precision can be improved further (e.g., AITZ).

\subsection{Analysis}
\subsubsection{Dynamic Evaluation of TSR}
Previous benchmark evaluations have been based on static analysis, which limits the autonomous planning and generality of the GUI agent. Thus, we also report the real-world TSR on mobile devices. As shown in Table~\ref{efficiency}, the baselines only achieve TSR of 4\% and 26\%. Given that the TSR of GPT-4o is 36\%, we see that OS-Kairos is approaching this upper limit. When OS-Kairos${_\text{human}}$ is assisted by human intervention, the TSR increases from 32\% to 70\%, indicating adaptive interaction is an effective paradigm for real-world GUI agents.

\begin{table}[t]
    \centering
    \renewcommand{\arraystretch}{1.2}
    \resizebox{\linewidth}{!}{
    \begin{tabular}{lcccc}
    \toprule 
    \textbf{Models}  & \textbf{Human Steps} & \textbf{Actual Steps} & \textbf{RE (\%)}$\uparrow$ & \textbf{TSR (\%)}$\uparrow$ \\ \midrule
    GPT-4o & 229 & 302 & 75.83 & 36.00 \\
    Qwen2-VL-7B  & 229 & 397 & 57.68 & 4.00 \\
    OS-Atlas-Pro-7B  & 229 & 359 & 63.79 & 26.00 \\ \midrule
    OS-Kairos$_{\text{GPT-4o}}$  & 229 & \textbf{245} & \textbf{93.47} & 32.00 \\
    OS-Kairos$_{\text{human}}$  &229& 265&	86.42	& \textbf{70.00} \\ \bottomrule
    \end{tabular}}
    \caption{Analysis of efficiency and dynamic TSR.}
    \label{efficiency}
\end{table}

\subsubsection{Efficiency Evaluation}
Table~\ref{efficiency} reports the efficiency in a real-world environment. First, we count the optimal number of human steps on 50 instructions, about 429 steps. Next, we evaluate the actual step counts for baseline and OS-Kairos, respectively. Notably, the model max steps are set to 10. We observe that the baseline models tend to over-execute when faced with a complex step. In contrast, OS-Kairos more closely resembles human manipulation of a GUI, achieving 86.42\% and 93.47\% in RE. 

\subsubsection{Comparing Prompt-based Interaction}
Table~\ref{HCI intervention} presents a comparison of OS-Kairos with prompt-based interactive models. We see that the interactive mechanism of OS-Kairos outperforms the prompt-based paradigm, particularly surpassing the prompt-based OS-Atlas-Pro-7B in terms of HSR. Despite the strong grounding capabilities of GPT-4o and GLM-4V-Plus, API-based agents present instability, resulting in over-execution and sub-optimal performance. Among open-source GUI agents, Qwen2-VL-7B performs more consistently than OS-Atlas-Pro-7B, because prompt-based interactive severely disrupts the latter's instruction-following ability. 
\begin{table}[t]
    \centering
    \LARGE
    \renewcommand{\arraystretch}{1.2}
    \resizebox{\linewidth}{!}{
    \begin{tabular}{lcccccc}
    \toprule 
    \textbf{Models} & \textbf{Interactive} & \textbf{Type (\%)}$\uparrow$ & \textbf{SR (\%)}$\uparrow$ & \textbf{TSR (\%)}$\uparrow$ & \textbf{HSR (\%)}$\uparrow$\\ \midrule
    GPT-4o & Prompt &88.80 &79.25 &46.58&/ \\
    GLM-4V-Plus & Prompt &88.34 &79.03 &47.83 &/ \\ 
    Qwen2-VL-7B & Prompt &76.42 &38.44 &25.47 &/  \\
    OS-Atlas-Pro-7B & Prompt &59.02 &95.67 &9.94 &0.00
    \\ \midrule
    OS-Kairos & FT & \textbf{99.88} & \textbf{95.90} &\textbf{88.20}&\textbf{86.87} \\ \bottomrule
    \end{tabular}}
    \caption{Analysis of interactive paradigms vs. prompt-based baseline in complex scenarios.}
    \label{HCI intervention}
\end{table}

\subsubsection{Ablation of Critic Models}\label{critic models}
As the advanced judgment capabilities of GPT-4o~\cite{chenmllm}, we utilize it as the critic model in the collaborative probing framework. To analyze the impact of the critic model on OS-Kairos confidence integration and GUI adaptive interaction, we select Qwen-VL-MAX as an alternative. Table~\ref{tab_critic_models} presents the adaptive interaction performance of OS-Kairos across different critic models. The results show that the scoring quality of GPT-4o significantly outperforms Qwen-VL-Max, with an HSR of 86.87\% compared to 57.63\%. In addition, the precision of the intervention decreases by 4.59\% in the autonomous steps and 9.25\% in the complex steps. Although GUI performance is similar, Qwen-VL-Max leads to more frequent interventions with OS-Kairos.
\begin{table*}[t]
    \centering
    \footnotesize
    \begin{tabular}{ccccccc}
      \toprule
      \textbf{Models} & \textbf{Type (\%)}$\uparrow$ & \textbf{SR (\%)}$\uparrow$ & \textbf{TSR (\%)}$\uparrow$ & \textbf{HSR (\%)$\uparrow$} & \textbf{IP(\%)}$\uparrow$& \textbf{AP(\%)}$\uparrow$ \\ \midrule
      GPT-4o & 98.49 & \textbf{96.36} & \textbf{87.71} & \textbf{86.87} & \textbf{70.75} & \textbf{96.44} \\
      Qwen-VL-MAX & \textbf{99.65} & 96.01 & 85.71 & 57.63 & 61.50 & 91.55 \\
      \bottomrule
    \end{tabular}
    \caption{Ablation of critic models.}
    \label{tab_critic_models}
\end{table*}

\begin{figure}[t]
    \centering
    \includegraphics[width=1\linewidth]{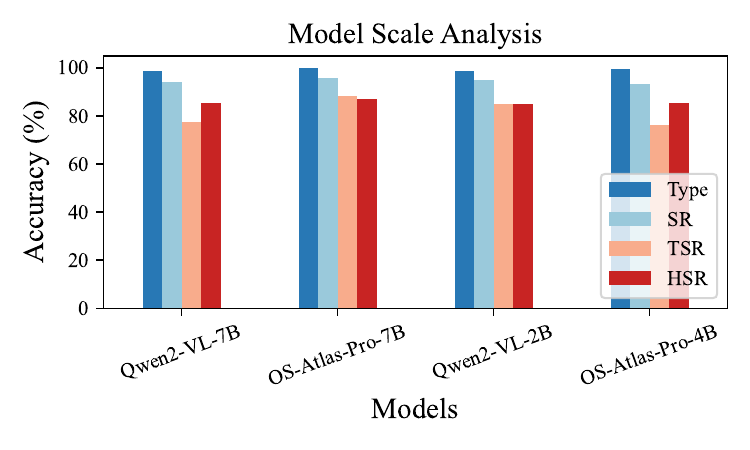}
    \vspace{-0.8cm}
    \caption{Generality of OS-Kairos across model scale.}
    \label{model_scale_law}
    \vspace{-0.2cm}
\end{figure}

\subsubsection{General Effectiveness across Scales}

\paragraph{Model Scale.}
Although the dynamic detection framework is built on the OS-Atlas-Pro-7B backbone, confidence scores and actions generated are supposed to be downwardly compatible. In other words, weaker models can be enhanced through data distillation and confidence scoring integration. Figure~\ref{model_scale_law} shows that OS-Kairos can be successfully generalized to the 2B$\sim$7B model. First, Type and SR are effective, guaranteeing a TSR of 76.40\% on the Qwen2-VL-7B, 77.64\% on OS-Atlas-Pro-4B, and 85.09\% on Qwen2-VL-2B. Thus, the combination of confidence scoring and data distillation will enhance weak models, thus satisfying the deployment in resource-constrained environments.

\paragraph{Data Scale.} To evaluate the effect of data scaling on confidence scoring integration, we divide the trajectories from the probing framework into different scales for training and test data. As shown in Table~\ref{data_scaling_law}, OS-Kairos is stable in Type and SR scores across data scales. Benefiting from its high HSR, OS-Kairos's TSR accuracy reaches 76.19\%$\sim$88.20\%, proving that the integration of confidence scoring into OS-Kairos requires only a small number of probing data at a significantly lower cost than fine-tuning the GUI agent.

\begin{table}[t]
    \centering
    \renewcommand{\arraystretch}{1.1}
    \resizebox{\linewidth}{!}{
    \begin{tabular}{cccccc}
    \toprule 
    \textbf{Data Scaling}  & \textbf{Type (\%)}$\uparrow$ & \textbf{SR (\%)}$\uparrow$ & \textbf{TSR (\%)}$\uparrow$ & \textbf{HSR (\%)}$\uparrow$\\ \midrule
    9:1  &99.25	&92.21	&76.19&	84.67 \\
    8:2  &\textbf{99.88}	&\textbf{95.90}	&\textbf{88.20}	&\textbf{86.87} \\ 
    7:3  &99.46	&94.16&	83.94&	84.79  \\
    6:4  &99.41	&94.05&	78.30	&84.47 \\ \bottomrule
    \end{tabular}}
    \caption{Varying data scale in confidence scoring.}
    \label{data_scaling_law}
\end{table}

\begin{figure}[t]
    \centering
    \includegraphics[width=1\linewidth]{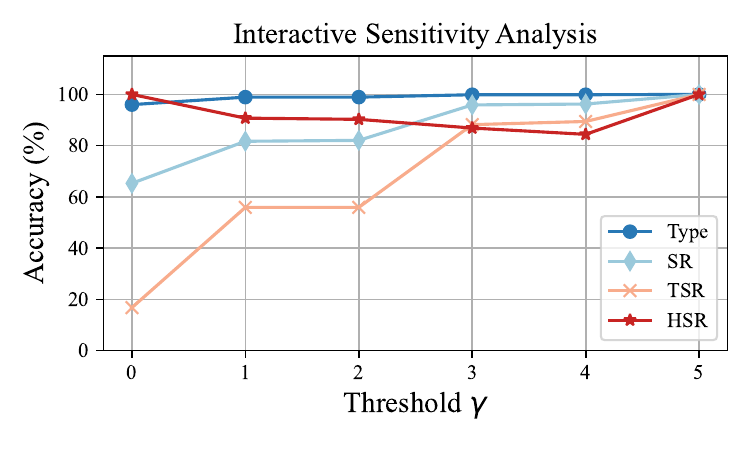}
    \vspace{-0.8cm}
    \caption{Threshold impact on interactive sensitivity.}
    \label{HSR sentivity}
\end{figure}

\begin{table}[t]
    \centering
    \resizebox{\linewidth}{!}{
    \begin{tabular}{ccccc}
      \toprule  
      \textbf{Threshold} & \textbf{Dataset} & \textbf{OS-Karios} & \textbf{SR}$\uparrow$ & \textbf{TSR}$\uparrow$ \\ 
      \midrule
      $\gamma=4$ & 28.60 & 37.28 & 95.90$_{\textcolor{red}{34.54\uparrow}}$ & 88.20$_{\textcolor{red}{73.91\uparrow}}$ \\
      $\gamma=2$ & 16.54 & 19.01 & 81.69$_{\textcolor{red}{20.33\uparrow}}$ & 55.28$_{\textcolor{red}{40.99\uparrow}}$ \\
      \bottomrule
    \end{tabular}}
    \caption{Comparison of human-intervention rates on the dataset (Column 2) and the actual OS-Kairos (Column 3) across different thresholds.}
    \vspace{-0.3cm}
    \label{tab7}
\end{table}

\subsubsection{Interactive Sensitivity}
OS-Kairos use a threshold to achieve adaptive interaction. To analyze the adaptive interaction sensitivity of OS-Kairos, we ablate threshold $\gamma$ from 0 to 5. In Figure~\ref{HSR sentivity}, TSR and SR increase with the rise in interactive sensitivity, indicating that human intervention enhances the effectiveness of GUI agents in complex scenarios. The HSR and Type accuracy remain stable across different thresholds, indicating that OS-Kairos can effectively identify complex steps, especially in coordinates and input scenarios, alleviating over-execution of the GUI agent. Furthermore, Table~\ref{tab7} reports the comparison results of the human intervention rate. In the defined complex scenario datasets, OS-Kairos exhibits a human-intervention rate similar to the real scenario. For example, when $\gamma=2$, OS-Kairos requires only 19\% intervention (0.81 steps for each instruction) to achieve results comparable to the fine-tuned model. The ablation study of adaptive interaction shows that OS-Kairos is more flexible (See Appendix~\ref{adpative}).

\section{Conclusion}
This study identifies a key challenge of over-execution in GUI agents, which poses substantial risks in complex scenarios, such as those involving ambiguous user
instructions, unexpected interruptions, and environmental hijacks.
To address the challenge, we introduce OS-Kairos, an adaptive GUI agent capable of predicting confidence levels at each step and efficiently deciding whether to act autonomously or seek human intervention. 
% We identify complex scenarios that impact TSR in GUI agents and provide a detailed theory analysis. 
Concretely, we propose a collaborative probing framework for annotating confidence scores at each interaction step. By integrating confidence scoring, OS-Kairos outperforms previous GUI agents and API-based models, with improved effectiveness, scalability, generality, and efficiency.

\section*{Limitations}
We acknowledge two primary limitations in our study. First, we only sampled instructions from three typical complex scenarios, as our focus was to investigate why existing GUI agents struggle with TSR and generate action confidence scores without loss of generality. Notably, we demonstrate the effectiveness of OS-Kairos on the AITZ and Meta-GUI benchmarks, which provide additional diverse instructions for complex scenarios. Besides, the generalization capabilities of OS-Kairos can mitigate these limitations. Second, experiments were focused on our probing dataset and two benchmark datasets, highlighting the need for complex scene probing and confidence scoring integration. Given that confidence scoring relies on proprietary models and high-quality human sampling, we anticipate that future research will explore the optimization of our approach to confidence scoring and evaluate new benchmark datasets.

\section*{Ethics Statement}
This section presents the ethics statements in the following aspects: (i) Privacy. The probing instructions are sourced from publicly available datasets, human designers, and GPT-4o, covering 12 apps and 12 topics. Temporary accounts were used to register these apps, and the trajectories generated by our collaborative probing framework are available, ensuring that no personal data or personally identifiable information was collected. The two benchmarks employed also implemented safeguards to protect privacy~\cite{zhang2024android,sun2022meta}. Moreover, OS-Kairos, as an open-source GUI agent that does not rely on any external information and supports local deployment. (ii) System security. OS-Kairos follows the first principles thinking \citep{zhang2023you}, manipulates the GUI like a human being, and can initiate human intervention in scenarios involving system security to ensure safety. (iii) Potential social impacts. OS-Kairos can improve the effectiveness of GUI execution instructions. Unlike fully autonomous GUI agents, OS-Kairos will proactively request authorization and acquire personal information, thus reducing malicious abuse.

% \section*{Acknowledgements}
% This work is partially supported by the Joint Funds of the National Natural Science Foundation of China (U21B2020), National Natural Science Foundation of China (62406188), and Natural Science Foundation of Shanghai (24ZR1440300).
% \vspace{-0.5cm}
\bibliography{acl_latex}

\appendix

\section{Why GUI agents have poor TSR?}\label{proof}
In our pilot experiment, the TSR of autonomous GUI agents is significantly lower than interactive GUI agents. This difference is attributed to the poor exact match for SR, particularly for the $\mathtt{CLICK}$ and $\mathtt{TYPE}$ actions. In contrast, interactive GUI agents can mitigate this limitation through human intervention. Intuitively, we consider the impact of exact matching on trajectory steps to be exponential. Formally, for a trajectory with $k$ steps, the probability that instruction $\tau_i$ can be completed is:
\begin{equation}
    \text{TSR}_{\tau_i} = \prod_{j=1}^k \text{SR}_{j}, s.t., SR_j \sim \beta(u, l).
\end{equation}
Herein, we assume that SR$_j$ follows $\beta$ distribution~\cite{mcdonald1995generalization}. $u$ and $l$ are hyperparameters that control the distribution of SR. The expectation $\mathbb{E}[\text{SR}_{\tau_i, j}] = \frac{u}{u+l}$ and variance $\text{Var}(\text{SR}_{\tau_i, j}) = \frac{ul}{(u+l)^2(u+l+1)}$. Additionally, the expectation $\mu=\mathbb{E}[\ln(\text{SR}_{\tau_i, j})] = \psi(u) - \psi(u+l)$, and the variance $\sigma^2=\text{Var}[\ln(\text{SR}_{\tau_i, j})] = \psi'(u) - \psi'(u+l)$, where $\psi(\cdot)$ and $\psi'(\cdot)$ denote the digamma and trigamma functions, respectively. The TSR$_{\tau_i}$ follows a normal distribution:
\begin{equation}
\ln(TSR_{{\tau_i}}) \sim \mathcal{N} \left( k \cdot \mu \right. \left. , \, k\cdot \sigma^2 \right),
\end{equation}
then, 
\begin{equation}
\begin{aligned}
    TSR_{{\tau_i}} &\sim \text{LogNormal}(\exp^{k\mu+\frac{k\sigma^2}{2}}, \,  \\
    &\exp^{2k\mu+k\sigma^2}(\exp^{k\sigma^2}-1)).
\end{aligned}
\end{equation}

\begin{figure}[ht]
    \centering
    \includegraphics[width=1\linewidth]{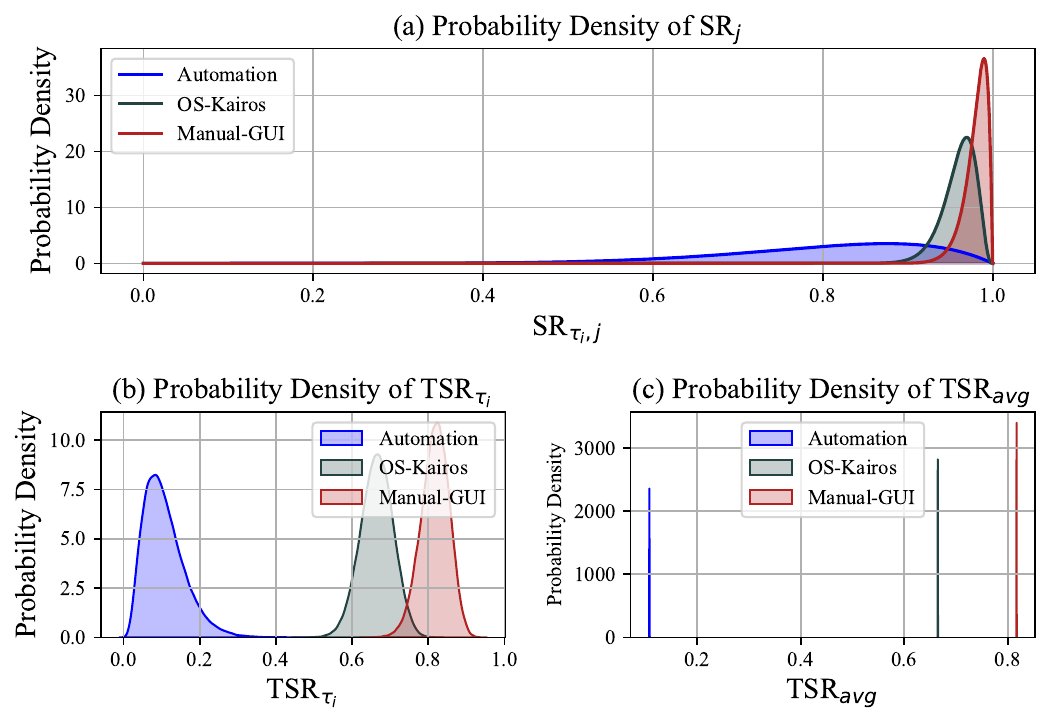}
    \caption{Illustration of the probability density of SR$_{\tau_i, j}$, TSR${\tau_i}$ for a trajectory, and TSR$_{\text{avg}}$ for $N$ trajectories.}
    \label{proof_fig_1}
\end{figure}

Considering the boundedness of $k$, we utilize Monte Carlo simulations~\cite{couto2013monte} to estimate the \textit{TSR$_{\tau_i}$} probability distribution. As shown in Figure~\ref{proof_fig_1}(a), we assume that SR$_{\text{auto}}$ exhibits high variance in the beta distribution, due to the effect of step complexity. The SR$_{\text{manual}}$, determined by the human intervention executed at each step, lies within the right-interval and represents the upper limit of the GUI agent’s capability. 
In this study, we aim to adaptively interact to bring OS-Kairos closer to this upper bound. As shown in Figure~\ref{proof_fig_1}(b), we observe that TSR$_{\tau_i, \text{auto}}$ is impacted by the complexity of the step, with the probability of TSR$_{\tau_i}$ falling below 20\%. In contrast, OS-Kairos can align with expectations and remains consistently close to the upper bound of performance.

When generalized to $N$ independent and identically distributed instructions, the average TSR satisfies:
\begin{equation}
    \text{TSR}_{avg} = \frac{1}{N} \sum_{i=1}^N \prod_{j=1}^k \text{SR}_{j}.
\end{equation}
According to center limit theory, \text{TSR$_{avg}$} also satisfies normal distribution:
\begin{equation}
\begin{aligned}
& TSR_{{avg}} \sim \mathcal{N}(\exp^{k\mu+\frac{k\sigma^2}{2}}, \, \\
& \exp^{2k\mu+k\sigma^2}(\exp^{k\sigma^2}-1)/N).
\end{aligned}
\end{equation}
As shown in Figure~\ref{proof_fig_1}(c), we observe the exponential effect of SR on TSR. In the autonomous mode, TSR$_{\text{avg}, \text{auto}}$ is nearly 0\%. In contrast, OS-Kairos and the fully interactive GUI agent both achieve success rates exceeding 60\%.

Subsequently, we further assume that the SR of single, complex, and interactive steps are $m, q, p$ respectively. When $\delta$ complex steps are available, TSR$_{\tau_i}$ satisfies:
\begin{equation}
    0 \approx m^\delta \cdot q^{k-\delta}  < \text{TSR}_{\tau_i} < m^\delta \cdot p^{k-\delta}  \approx p^k.
\end{equation}
\begin{figure}[ht]
    \centering
    \includegraphics[width=1\linewidth]{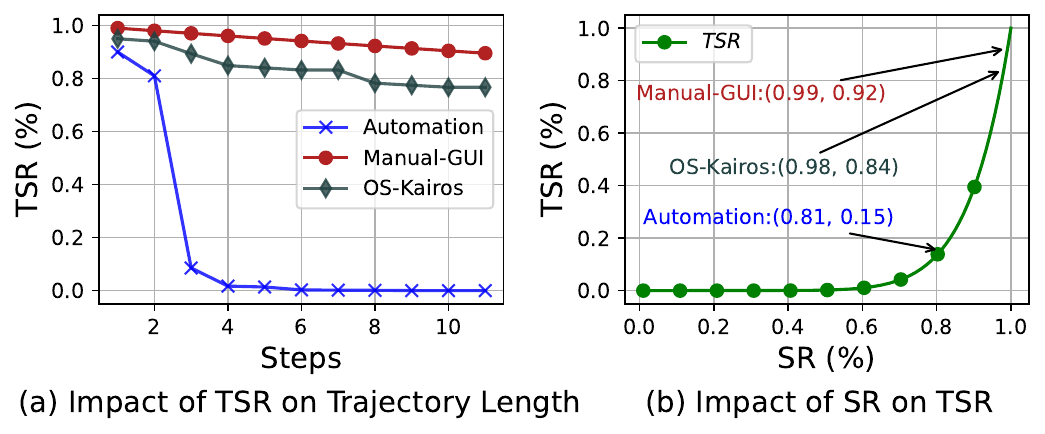}
    \caption{Illustration of the effect of SR on \textit{TSR$_{avg}$} for $N$ trajectories.}
    \label{proof_fig}
\end{figure}
As shown in Figure~\ref{proof_fig}(a), the SR effect on the TSR of a single trajectory is consistent with Figure~\ref{proof_fig_1}(a). In other words, once there are complex steps in the trajectory, the  TSR$_{\tau_i}$ will decrease significantly, while human intervention can jump such steps, thus remaining effective. 
Therefore, OS-Kairos aims to recognize such steps, seek human intervention, and thus exponentially enhance TSR, as shown in Figure~\ref{proof_fig}(b).

\section{Instruction Distribution}\label{distribution}
\begin{figure}[ht]
    \centering
    \includegraphics[width=1\linewidth]{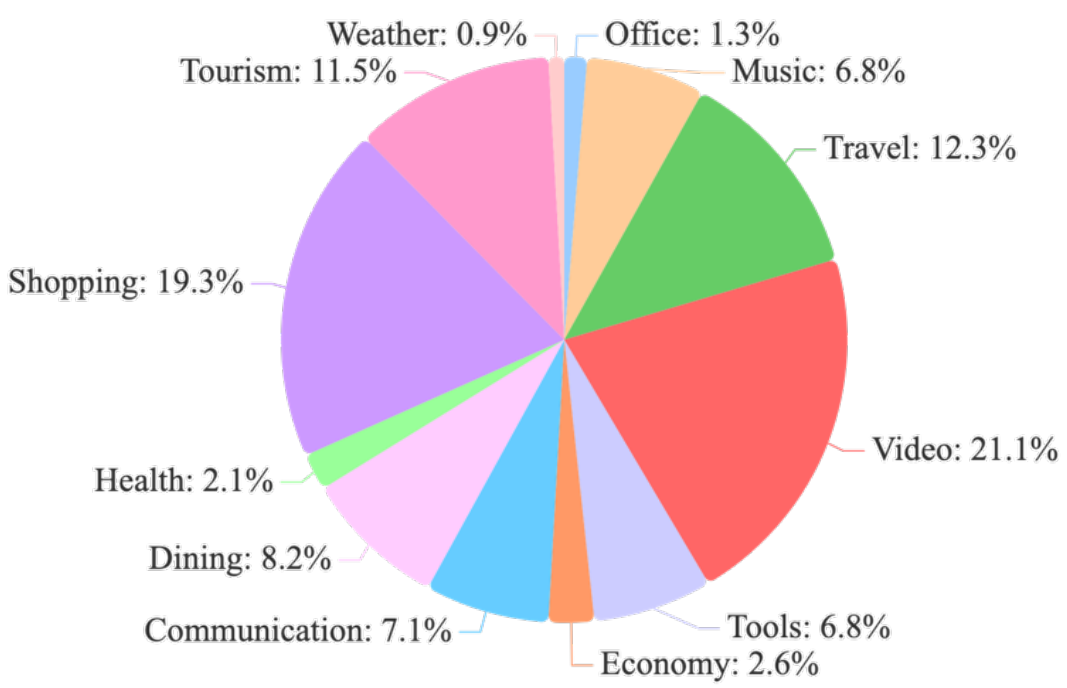}
    \caption{Subject distribution of instructions.}
    \label{statistics_}
\end{figure}

\begin{figure}[ht]
    \centering
    \includegraphics[width=1\linewidth]{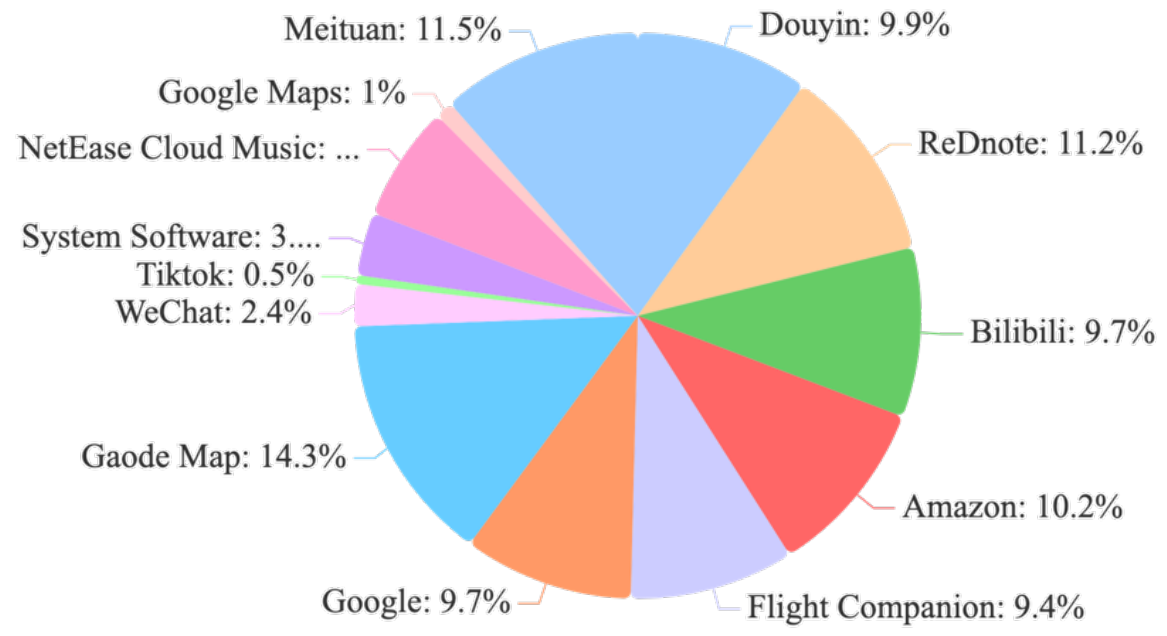}
    \caption{APP distribution of instructions.}
    \label{statistics_2}
\end{figure}

\begin{figure}[ht]
    \centering
    \includegraphics[width=1\linewidth]{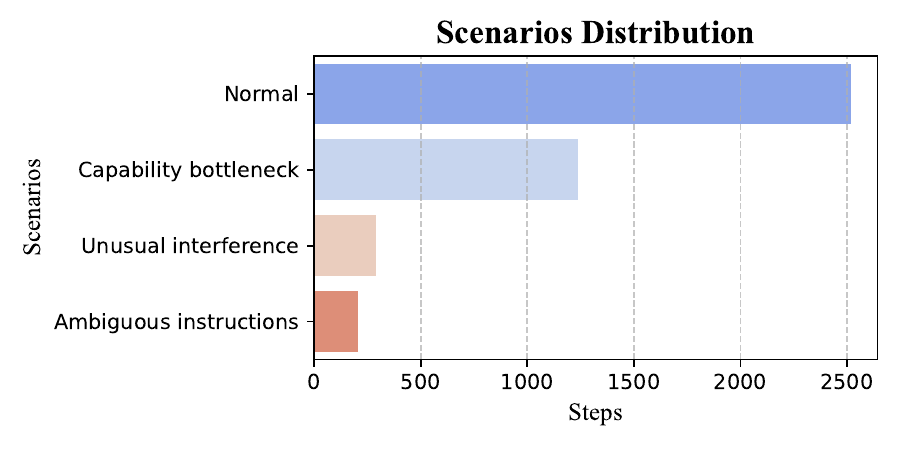}
    \caption{Distribution of steps in different scenarios.}
    \label{statistics}
\end{figure}

In our dynamic capability probing, we collect 1,000 instructions for three complex scenarios, covering 12 topics and 13 apps. The distributions of topics and apps are shown in Figure~\ref{statistics_} and Figure~\ref{statistics_2}. The distribution of scenarios is shown in Figure~\ref{statistics}.

\section{Detailed Experimental Setup}\label{Implementation Details}
\subsection{Prompt Templates} \label{Prompts}
\begin{figure}[ht]
    \centering
    \includegraphics[width=1\linewidth]{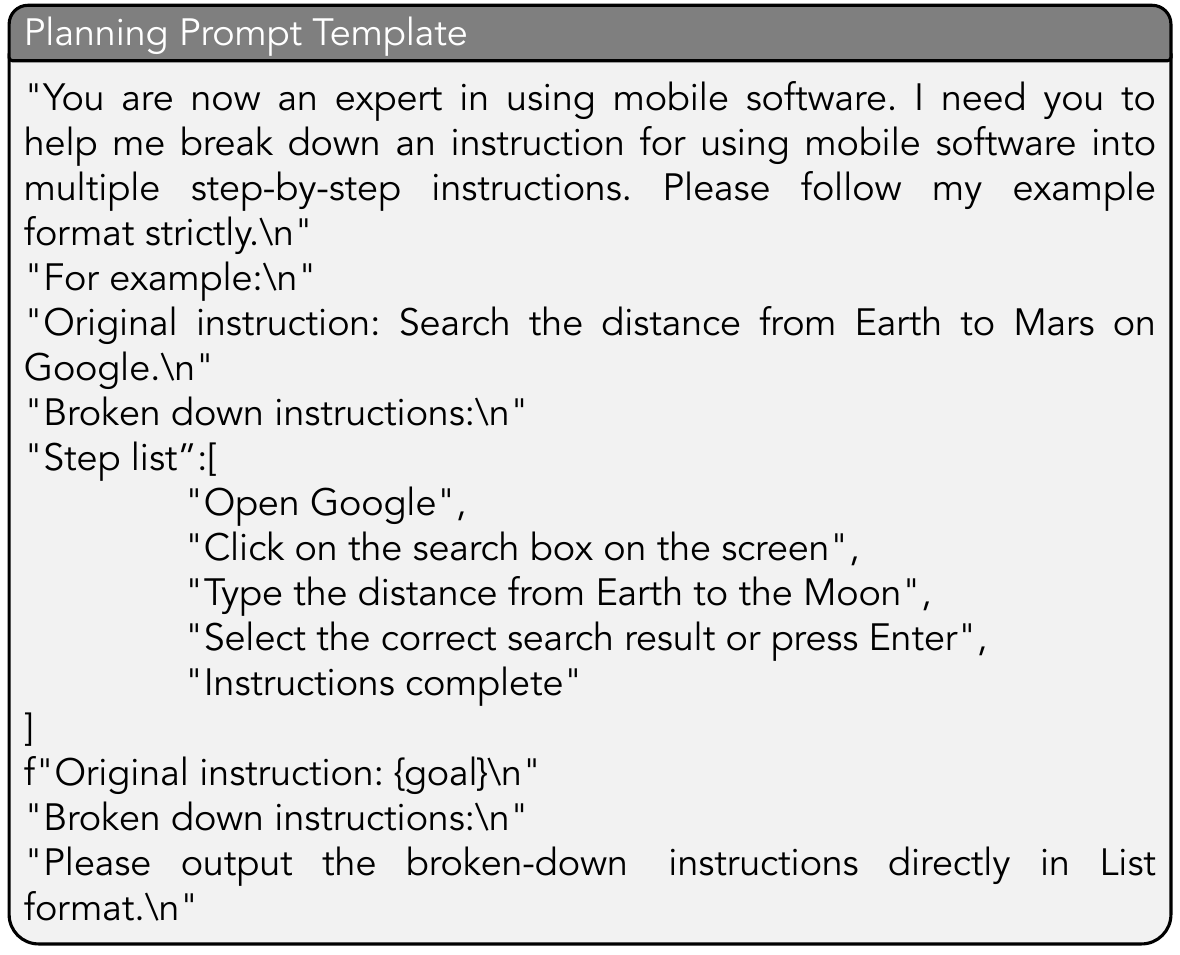}
    \caption{Prompts of the critic model for generating the planning schedule of the user instruction.}
    \label{step_list_prompt}
\end{figure}

Below are the prompt templates for designing OS-Kairos. In the collaborative probing framework, we design a planning prompt, action phase prompt, action prompt for probed GUI agent $\mathcal{F}_p$ and critic model $\mathcal{F}_c$, scoring prompt, and finishing judgment prompt. In the confidence-driven interaction phase, we only use the $\mathcal{F}_p$ action prompt to optimize and evaluate OS-Kairos. The pipeline controller fills these {{variables}} in them according to the context. 
\noindent\textbf{Planning Prompt Template.}
In the GUI capability probing framework, the critic model $\mathcal{F}_c$ generates the planning schedule of the user instruction based on the GPT-4o and the instruction planning prompt, as shown in Figure~\ref{step_list_prompt}.

\begin{figure}[ht]
    \centering
    \includegraphics[width=1\linewidth]{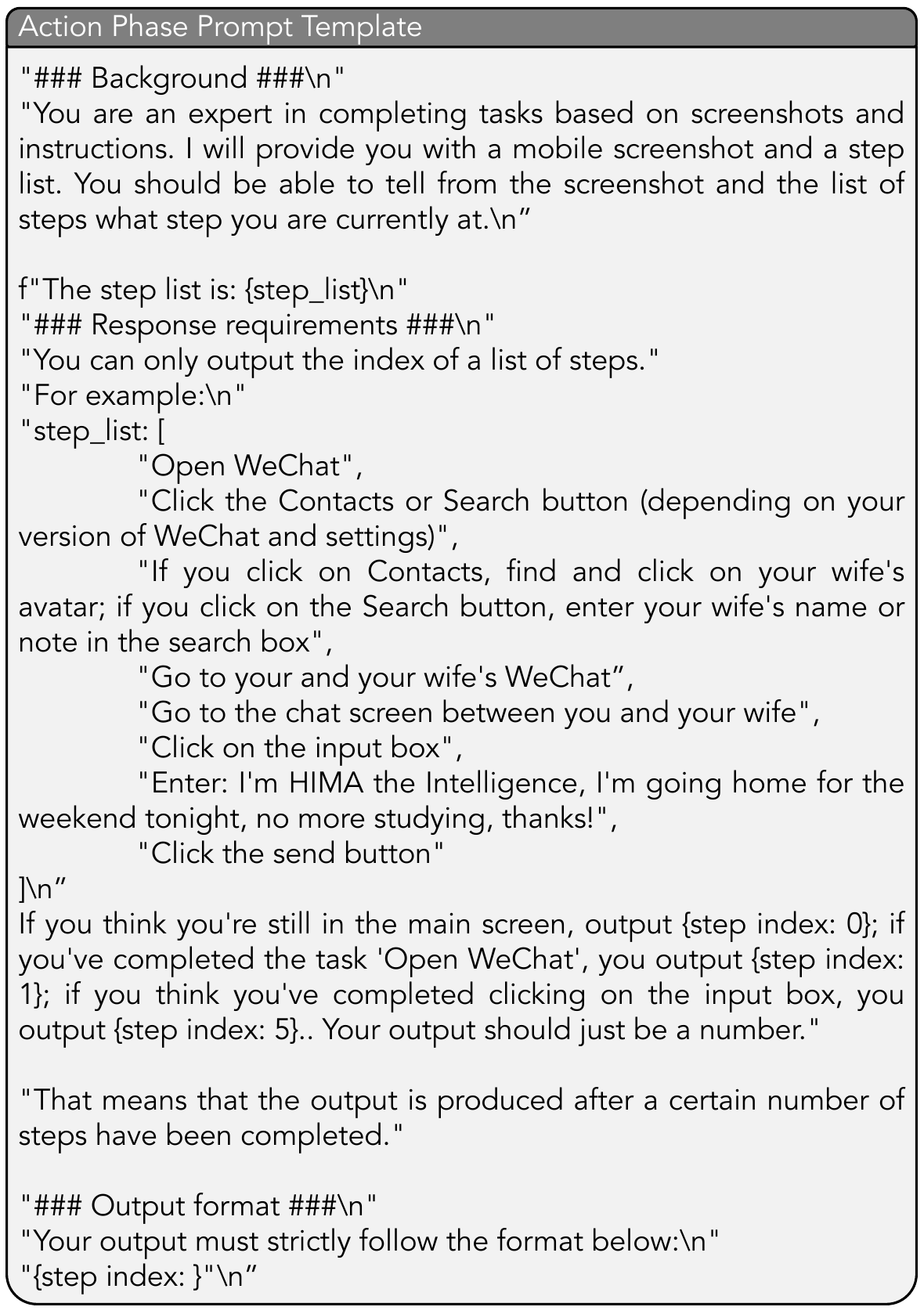}
    \caption{Prompts of the critic model for generating the action phase.}
    \label{action_phase}
\end{figure}

\noindent\textbf{Action Phase Prompt Template.}
In the collaborative probing framework, the critic model $\mathcal{F}_c$ determines the current step based on the GPT-4o and action phase prompts, as shown in Figure~\ref{action_phase}.

\begin{figure}[ht]
    \centering
    \includegraphics[width=1\linewidth]{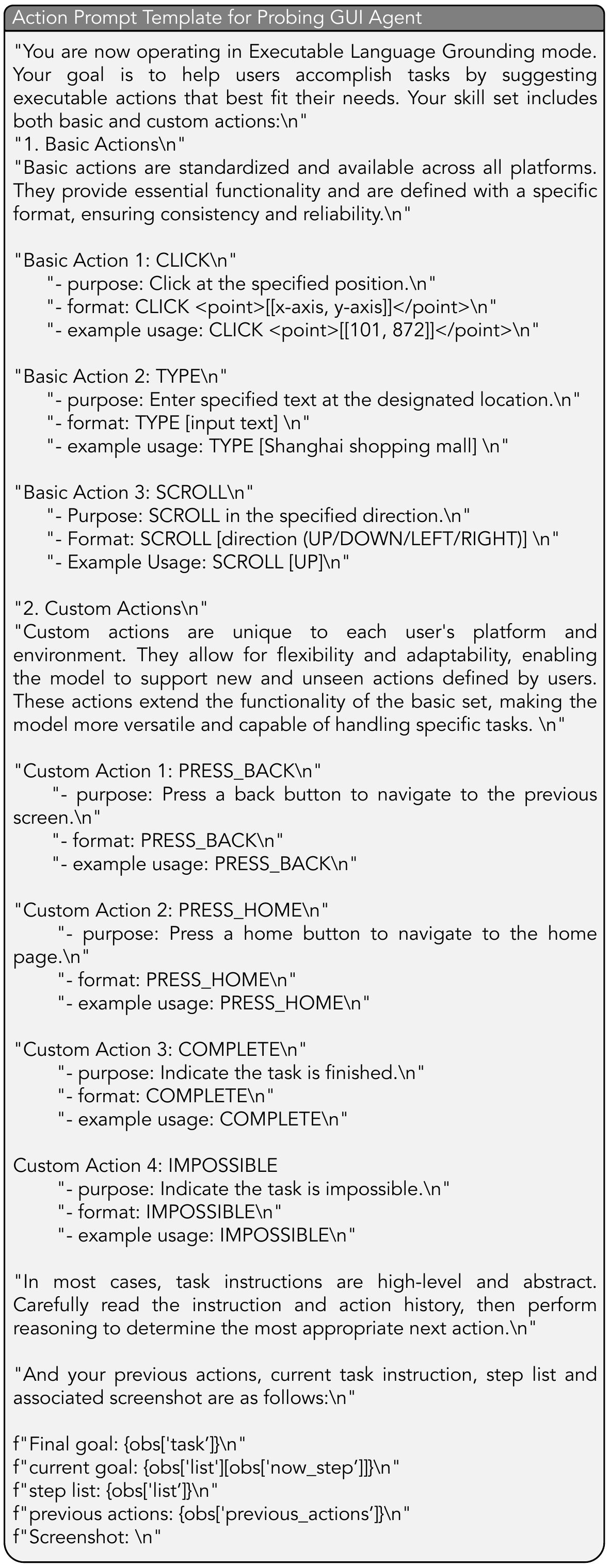}
    \caption{Prompt of the probed GUI agent for generating action.}
    \label{os_action}
\end{figure}

\begin{figure}[ht]
    \centering
    \includegraphics[width=1\linewidth]{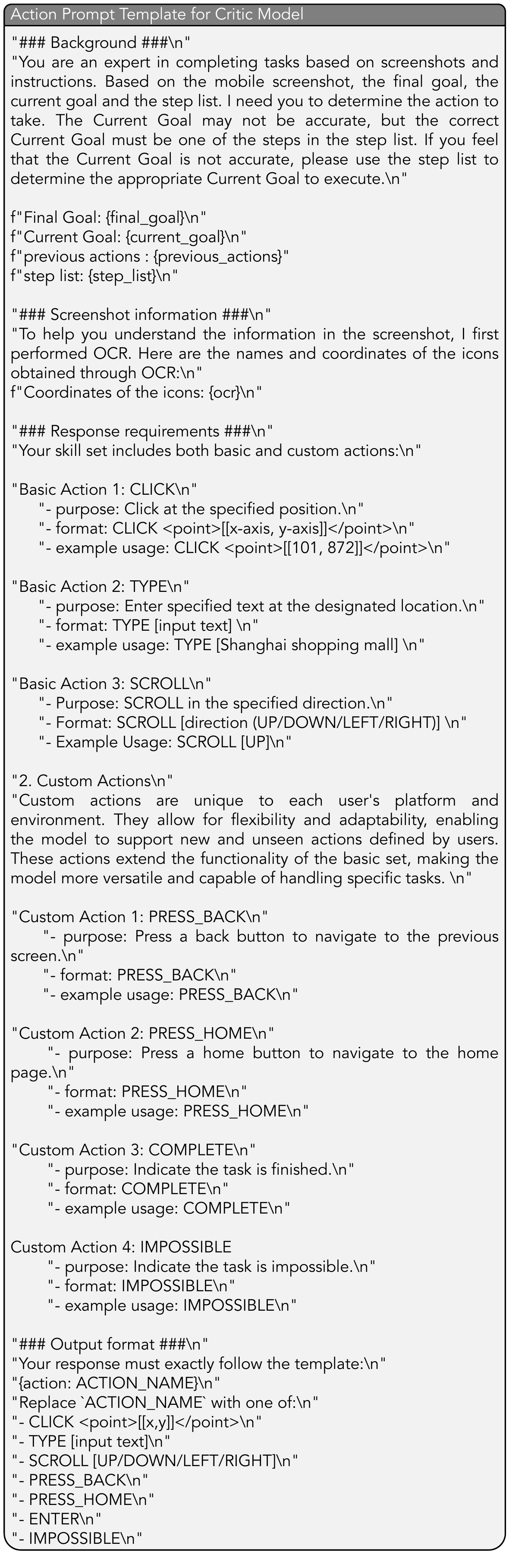}
    \caption{Prompt of the critic model for generating action.}
    \label{gpt_action}
\end{figure}

\noindent\textbf{Action Prompt Template.}
In the collaborative probing framework, we obtain the prediction of GUI agents using action prompts. The action prompts for the probed GUI agent $\mathcal{F}_p$ and critic model $\mathcal{F}_c$ are shown in Figure~\ref{os_action} and Figure~\ref{gpt_action}, respectively. Following~\cite{wu2024atlas, zhang2024android}, we define the actions set, which comprises 7 kinds of actions: $\mathtt{CLICK}$, $\mathtt{SCROLL}$, $\mathtt{TYPE}$, $\mathtt{PRESS\_BACK}$, $\mathtt{PRESS\_HOME}$, $\mathtt{COMPLETE}$, and $\mathtt{IMPOSSIBLE}$.

\begin{figure}[ht]
    \centering
    \includegraphics[width=1\linewidth]{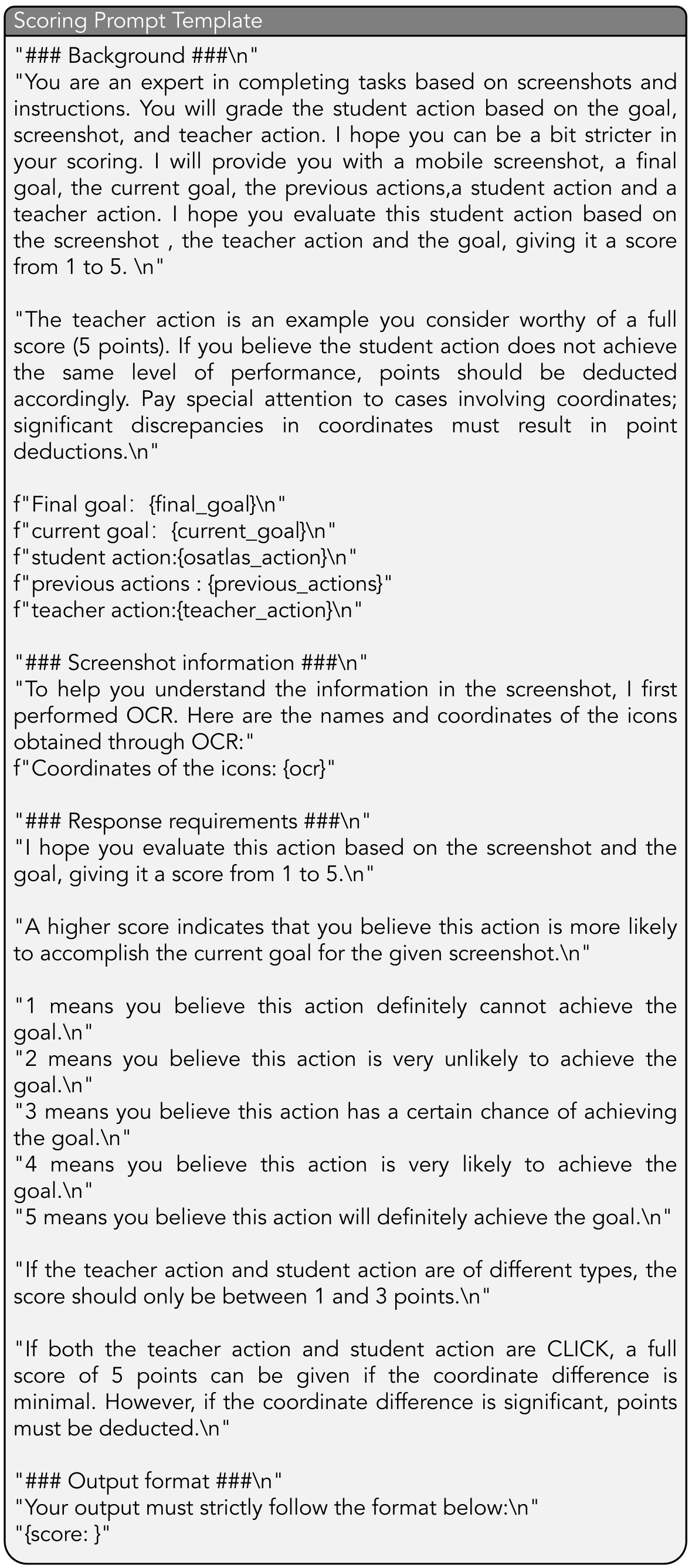}
    \caption{Prompt of the critic model for generating action score.}
    \label{score}
\end{figure}

\noindent\textbf{Scoring Prompt Template.}
In the collaborative probing framework, the critic model $\mathcal{F}_c$  generates the score for the action of $\mathcal{F}_p$ based on GPT-4o and scoring prompt, as shown in Figure~\ref{score}. 

\begin{figure}[ht]
    \centering
    \includegraphics[width=1\linewidth]{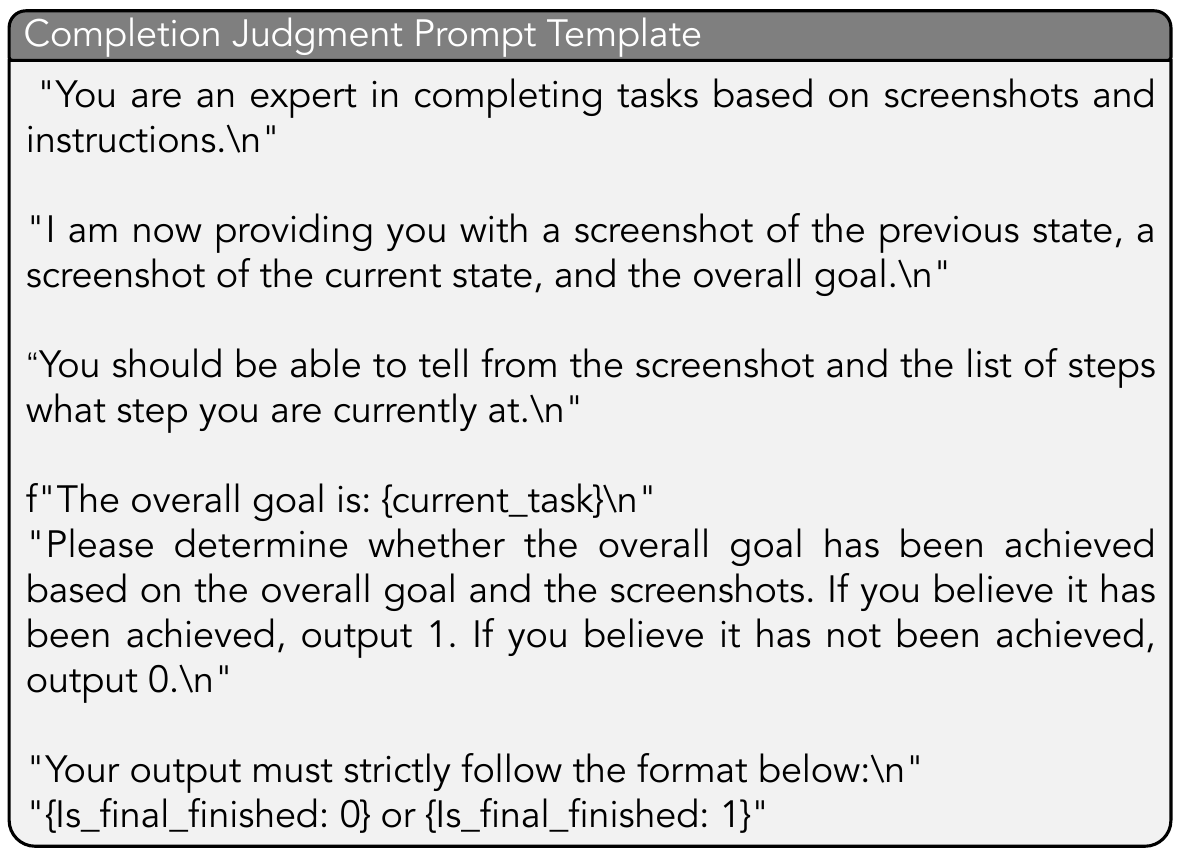}
    \caption{Prompt for the critic model to judge instruction completion.}
    \label{finshed}
\end{figure}

\noindent\textbf{Completion Judgment Prompt Template.}
In the collaborative probing framework, the critic model $\mathcal{F}_c$ exploits GPT-4o to judge whether the instruction is completed. The prompt is shown in Figure~\ref{finshed}. 

\subsection{Details of Datasets}\label{datasets_appendix}
We consider evaluating OS-Kairos on three customized complex scenarios and two benchmarks: AITZ~\cite{zhang2024android} and Meta-GUI~\cite{sun2022meta}. The statistics of the dataset are shown in the Table~\ref{dataset_statistics}.
\begin{itemize}[left=0em]
    \vspace{-0.15cm}
    \item \textbf{AITZ}~\cite{zhang2024android}: The first dataset to employ chain-of-action thought (CoAT) connects perception (of screen layouts and UI elements) with cognition (of action decision-making) to enhance the AITW benchmark. This dataset comprises 2,504 operation trajectories across 18.6K real-world intentions. Based on the application domain, AITZ is also divided into five subsets: General, Install, GoogleApps, Single, and WebShopping.
    \vspace{-0.15cm}
    \item \textbf{Meta-GUI}~\cite{sun2022meta}: task-oriented dialogue dataset is released for interactive GUI agent. These utterances cut a trajectory into several dialogue turns. Meta-GUI consists of 1,009 trajectories with 16.4K steps. The data diversity lies in 11 applications of 6 topics.
\end{itemize}

\begin{table}[t]
    \centering
    \small
    \begin{tabular}{lccc}
     \toprule
     \textbf{OS-Kairos} & \textbf{Trajectory} & \textbf{Screen} & \textbf{Goal} \\ \midrule
      Train & 800 & 4078 & 759  \\ 
      Test & 200 & 1054 & 198 \\ \midrule
      \textbf{AITZ} & \textbf{Trajectory} & \textbf{Screen} & \textbf{Goal} \\ \midrule
      General & 479 & 3607 & 479 \\
      Install & 420 & 3627 & 420 \\ 
      Google Apps & 242 & 1889 & 242 \\
      Single & 844 & 2594 & 844 \\
      Web Shopping & 519 & 6926 & 519 \\ \midrule
      \textbf{Meta-GUI} & \textbf{Trajectory} & \textbf{Screen} & \textbf{Goal} \\ \midrule
      Train & 897 & 14539 & 2286 \\
      Test  & 116 & 1923 & 336 \\ \bottomrule
    \end{tabular}
    \caption{Dataset statistics.}
    \label{dataset_statistics}
\end{table}

\subsection{Details of Evaluation Metrics}\label{metrics_appendix}
To ensure fair comparison across all baseline methods, we standardize the evaluation metrics for each action. We define the SR metrics for the three complex actions as follows: 
\begin{itemize}[left=0em]
    \item \textbf{$\mathtt{CLICK}$:} GUI agent predictions are considered correct if and only if both action types and position coordinates <x, y>. Following~\cite{zhang2023you}, we measure performance by calculating the distance between the predicted and ground truth coordinates. We consider the coordinates to be correct if the distance between the coordinates and the ground truth is within 14\% of the screen width.
    \item \textbf{$\mathtt{TYPE}$:} GUI agent predictions are considered correct if and only if both action type and action content are correct.
    \item \textbf{$\mathtt{SCROLL}$:} GUI agent predictions are considered correct if and only if both action type and direction argument (i.e., UP, DOWN, LEFT, and RIGHT)  are correct.
\end{itemize}
Furthermore, Type measures the exact match score between the predicted action types (e.g., CLICK, SCROLL) and the ground truth. TSR requires that all steps in a trajectory be correctly executed. For HSR, we define four statistical metrics with the threshold $\gamma$: 
\begin{itemize}[left=0em]
    \item \textbf{True positive (TP)}: Neither the prediction confidence nor the ground truth exceeds the $\gamma$, i.e., the agent does not require and perform interactions.
    \item \textbf{False positive (FP)}: The prediction confidence is greater than the $\gamma$, but the ground truth does not, meaning the agent is not required, but interaction is performed.
    \item \textbf{True negative (TN)}: Both the prediction confidence and the ground truth exceed the $\gamma$, meaning the agent must also perform interaction.
    \item \textbf{False negative (FN)}: The prediction confidence is less than $\gamma$, but the ground truth is greater than $\gamma$, which means that the agent needs but does not perform interactions.
\end{itemize}
Hence, HSR can be calculated:
\begin{equation}
    \text{HSR} = \frac{TP+TN}{TP+TN+FP+FN}.
\end{equation}
In addition, IP calculates the accuracy of the intervention step where intervention is actually needed, while AP measures the accuracy of the autonomous step where autonomy is truly required. Hence, IP and AP can be calculated:
\begin{equation}
    \text{IP} = \frac{TN}{TN+FN},\, \text{AP} = \frac{TP}{TP+FP}.
\end{equation}

Following~\cite{wang2024mobile}, RE measures the relative efficiency of the GUI agent compared to the steps taken by humans. It demonstrates whether OS-Kairos can use the mobile device more efficiently.

\subsection{Usage of Existing Artifacts}
For API-based MLLMs, we access them directly via the official interface. For open-source MLLMs, we either download the model weights from Hugging Face\footnote{\url{https://huggingface.co/models}} or reproduce the model using the same training strategy. In our proposed OS-Kairos, the layout-parse pipeline of the collaborative probing framework is built upon Modelscope~\footnote{\url{https://modelscope.cn/home}}. Furthermore, we utilize $\mathtt{LLaMA}$-$\mathtt{Factory}$\footnote{\url{https://github.com/hiyouga/LLaMA-Factory}} to fine-tune the probed model on three datasets for confidence integration. Notably, the InternVL-based models are fine-tuned using $\mathtt{Xtuner}$\footnote{\url{https://github.com/InternLM/xtuner}}. All licenses for these packages permit their use for standard academic research purposes.

\subsection{Further Analysis}
\subsubsection{AITZ Benchmark}\label{aitz-f}
Table~\ref{aitz} presents a comparison of OS-Kairos with the baselines in the AITZ benchmark. In API-based MLLMs, although GPT-4o performs the best, it is nearly impossible to finish user instructions. Among the open-source GUI agents, OS-Atlas-Pro-7B outperforms the other baselines due to the adaptation of AITZ, but still exhibits low SR and cannot fully complete user instructions. In contrast, OS-Kairos achieves precise intervention in complex steps on top of OS-Atlas-Pro-7B, with significant improvements in actions and overall performance. As a result, OS-Kairos's TSR increased from 0\% to 24.51\%.
\begin{table*}[t]
    \centering
    \Large
    \renewcommand{\arraystretch}{1.3}
    \resizebox{\linewidth}{!}{
    \begin{tabular}{lccccccccccc}
    \toprule
     \multirow{2}{*}{\textbf{Models}} & \multirow{2}{*}{\textbf{API}} &  \multirow{2}{*}{\textbf{SCROLL}} &  \multirow{2}{*}{\textbf{PRESS}} & \multirow{2}{*}{\textbf{STOP}} & \multicolumn{2}{c}{\textbf{CLICK}}& \multicolumn{2}{c}{\textbf{TYPE}}  & \multicolumn{2}{c}{\textbf{Total}} & \multirow{2}{*}{\textbf{TSR}} \\ \cmidrule(lr){6-7} \cmidrule(lr){8-9} \cmidrule(lr){10-11}
     & & & &  & \textbf{Type (\%)} $\uparrow$ & \textbf{SR (\%)} $\uparrow$ & \textbf{Type (\%)} $\uparrow$ & \textbf{SR (\%)} $\uparrow$ & \textbf{Type (\%)} $\uparrow$ & \textbf{SR (\%)} $\uparrow$ \\ \midrule
     GPT-4o &  \textcolor[RGB]{34,139,34}{\ding{51}}&24.17&23.84& 0.00&63.80&27.71&35.20&16.00&58.32&22.69&0.00 \\     
     GLM-4V-Plus  & \textcolor[RGB]{34,139,34}{\ding{51}}&11.65 &7.28 &0.00 &79.15 &27.65 &43.80 &20.40 &68.95 &20.92 &0.00\\
     Qwen-VL-MAX  & \textcolor[RGB]{34,139,34}{\ding{51}} &7.89 &13.04 &10.2 & / &72.3 &/ &34.04 &/ &52.41 & / \\ \midrule
     CogAgent  & \textcolor[RGB]{178,34,34}{\ding{55}} &56.41 &48.30 &4.76 &79.90 &51.50 &67.40 &34.00 &65.86 &44.52 &/ \\
     Auto-UI  & \textcolor[RGB]{178,34,34}{\ding{55}} &74.88 &49.09 &60.12 &44.37 &12.72 &73.00 &67.80 &73.79 &34.46 & /\\
     Qwen2-VL-7B  & \textcolor[RGB]{178,34,34}{\ding{55}} &18.64 &21.19 &0.00 &71.05 &32.89 &82.80 &45.00 &66.28 &28.25 &0.00 \\
     OS-Atlas-Pro-7B  & \textcolor[RGB]{178,34,34}{\ding{55}} &27.40 &0.66 &5.16 &93.31 &34.87 &85.20 &27.40 &85.20 &33.66 &0.00 \\ \midrule
     OS-Kairos &\textcolor[RGB]{178,34,34}{\ding{55}}  & \textbf{91.17}$_{\textcolor[RGB]{178,34,34}{63.77\uparrow}}$ & \textbf{73.51}$_{\textcolor[RGB]{178,34,34}{72.85\uparrow}}$ & \textbf{91.65}$_{\textcolor[RGB]{178,34,34}{86.49\uparrow}}$ & \textbf{98.43}$_{\textcolor[RGB]{178,34,34}{5.12\uparrow}}$ & \textbf{89.46}$_{\textcolor[RGB]{178,34,34}{54.59\uparrow}}$ & \textbf{99.20}$_{\textcolor[RGB]{178,34,34}{14.00\uparrow}}$ & \textbf{72.80}$_{\textcolor[RGB]{178,34,34}{45.40\uparrow}}$ & \textbf{96.81}$_{\textcolor[RGB]{178,34,34}{11.61\uparrow}}$ & \textbf{87.54}$_{\textcolor[RGB]{178,34,34}{53.88\uparrow}}$ & \textbf{24.51}$_{\textcolor[RGB]{178,34,34}{24.51\uparrow}}$ \\
     \bottomrule    \end{tabular}}
    \caption{Comparison of OS-Kairos with baselines in the AITZ benchmark (zero-shot setting). We report the overall accuracy for Type, SR, and TSR, along with fine-grained accuracy for each action. Subscripts indicate absolute improvement over the OS-Atlas-Pro-7B, with the best result highlighted in \textbf{bold}.}
    \label{aitz}
\end{table*}

\subsubsection{Meta-GUI Benchmark}\label{meta-gui-f}

Meta-GUI benchmark dataset is an out-of-domain (OOD) task against probing models, which allows for probing more complex steps and generating the confidence level for each step. Table~\ref{meta-gui-b} presents the performance of OS-Kairos on the Meta-GUI benchmark compared to the baseline. First, API-based MLLMs exhibit lower SR (17.19\% to 32.72\%) and Type (54.74\% to 69.85\%), which can be attributed to over-execution on complex steps such as $\mathtt{SCROLL}$. Hence, Qwen-VL-MAX only achieves a TSR of 15.42\%, while GLM-4V-Plus performs weakly, with only 1.67\% TSR. In addition, three open-source GUI agents such as OS-Atlas-Pro-7B are even less effective, as they cannot adapt to OOD instructions. In contrast, OS-Kairos achieves the accuracies of 98.49\% in Type, 96.36\% in SR and 87.71\% in TSR, respectively. Similarly, the fine-grained Type and SR outperform the baseline methods. 

\begin{table*}[ht]
    \centering
    \Large
    \renewcommand{\arraystretch}{1.3}
    \resizebox{\linewidth}{!}{
    \begin{tabular}{lccccccccccc}
    \toprule
     \multirow{2}{*}{\textbf{Models}} & \multirow{2}{*}{\textbf{API}} &  \multirow{2}{*}{\textbf{SCROLL}} &  \multirow{2}{*}{\textbf{PRESS}} & \multirow{2}{*}{\textbf{STOP}} & \multicolumn{2}{c}{\textbf{CLICK}}& \multicolumn{2}{c}{\textbf{TYPE}}  & \multicolumn{2}{c}{\textbf{Total}} & \multirow{2}{*}{\textbf{TSR}} \\ \cmidrule(lr){6-7} \cmidrule(lr){8-9} \cmidrule(lr){10-11}
     & & & &  & \textbf{Type (\%)} $\uparrow$ & \textbf{SR (\%)} $\uparrow$ & \textbf{Type (\%)} $\uparrow$ & \textbf{SR (\%)} $\uparrow$ & \textbf{Type (\%)} $\uparrow$ & \textbf{SR (\%)} $\uparrow$ \\ \midrule
     GPT-4o  & \textcolor[RGB]{34,139,34}{\ding{51}}&33.97 &25.00 &12.79 &94.12 &42.30 &66.47 &28.14 &69.85 &32.72 &6.67\\
     GLM-4V-Plus & \textcolor[RGB]{34,139,34}{\ding{51}}&0.00 &0.00 &1.06 &95.45 &26.53 &38.01 &22.81 &65.05 &17.19 &1.67\\
     Qwen-VL-MAX  & \textcolor[RGB]{34,139,34}{\ding{51}}&14.74 &40.91 &1.91 &70.87 &37.85 &74.85 &45.03 &54.74 &27.86 &15.42  \\ \midrule
     Auto-UI  & \textcolor[RGB]{178,34,34}{\ding{55}} &60.90 &0.00 &0.00 &26.90 &2.69 &0.00 &0.00 &20.02 &6.45 &0.00 \\
     Qwen2-VL-7B & \textcolor[RGB]{178,34,34}{\ding{55}} &0.00 &0.00 &0.43 &51.54 &2.33 &38.01 &18.13 &35.90 &3.04 &0.21 \\
     OS-Atlas-Pro-7B  & \textcolor[RGB]{178,34,34}{\ding{55}} &16.03 &0.00 &0.00 &94.53 &37.29 &60.23 &15.20 &66.09 &23.59 &0.42 \\ \midrule
     OS-Kairos  & \textcolor[RGB]{178,34,34}{\ding{55}} & \textbf{99.36}$_{\textcolor[RGB]{178,34,34}{83.33\uparrow}}$ & \textbf{100.0}$_{\textcolor[RGB]{178,34,34}{100.0\uparrow}}$ & \textbf{94.73}$_{\textcolor[RGB]{178,34,34}{94.73\uparrow}}$ & \textbf{99.81}$_{\textcolor[RGB]{178,34,34}{5.28\uparrow}}$ & \textbf{96.66}$_{\textcolor[RGB]{178,34,34}{59.37\uparrow}}$ & \textbf{98.83}$_{\textcolor[RGB]{178,34,34}{38.60\uparrow}}$ & \textbf{95.32}$_{\textcolor[RGB]{178,34,34}{80.12\uparrow}}$ & \textbf{98.49}$_{\textcolor[RGB]{178,34,34}{32.40\uparrow}}$ & \textbf{96.36}$_{\textcolor[RGB]{178,34,34}{72.77\uparrow}}$ & \textbf{87.71}$_{\textcolor[RGB]{178,34,34}{87.29\uparrow}}$ \\ \bottomrule
    \end{tabular}}
    \caption{Comparison of OS-Kairos with baselines in the Meta-GUI benchmark (zero-shot setting). We report the overall accuracy for Type, SR, and TSR, along with fine-grained accuracy for each action. Subscripts indicate absolute improvement over the OS-Atlas-Pro-7B, with the best result highlighted in \textbf{bold}.}
    \label{meta-gui-b}
\end{table*}

\subsubsection{Generality Evaluation of OS-Kairos}\label{ood_section}
\begin{figure*}[ht]
    \centering
    \includegraphics[width=1\linewidth]{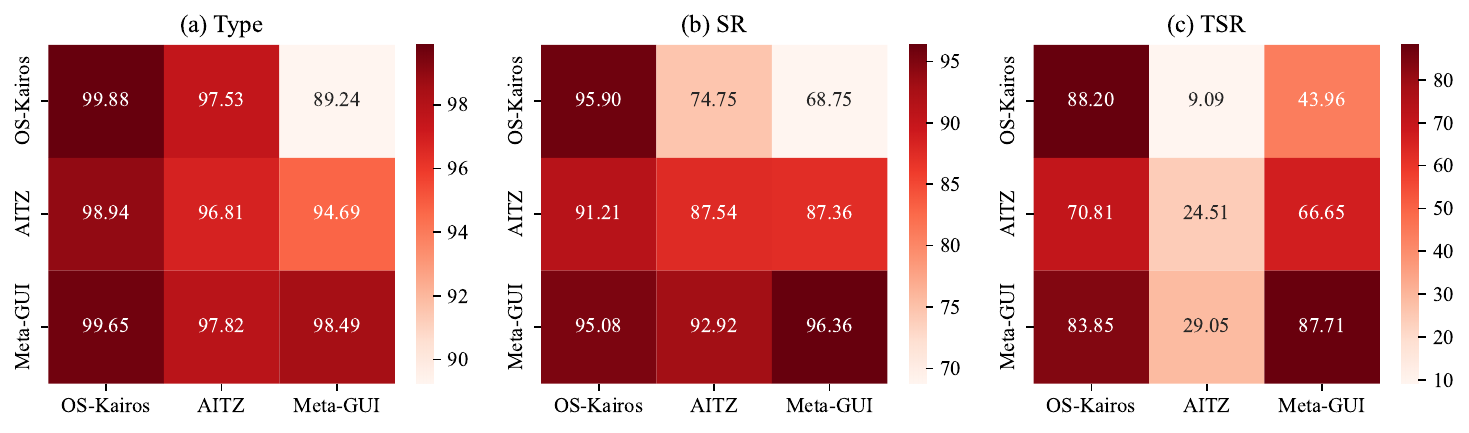}
    \caption{Generality analysis of OS-Kairos for adaptive interaction from original dataset to target datasets.}
    \label{ood}
\end{figure*}
OS-Kairos outperforms the baseline model across three datasets due to the integration of the confidence scoring. To verify the generality of adaptive interaction, we train OS-Kairos on each of the three datasets and then test it on the other two. The evaluation results are presented in Figure~\ref{ood}. We observe that OS-Kairos is able to achieve a decent performance, though the domains vary. Compared to the main results, it significantly outperforms the baseline in the zero-shot setting across three datasets, particularly in SR and TSR metrics. Also, its generalization performance is comparable to that of models fine-tuned directly on the target dataset. We also note that the more complex the dataset on which confidence scoring is integrated, the better the generalization of OS-Kairos. For example, OS-Kairos exhibits the best generalization with confidence scoring integration on the Meta-GUI dataset (29.05\% vs. 21.74\% in the AITZ benchmark, and 83.85\% vs. 88.20\% in the OS-Kairos dataset).

\subsubsection{Ablation of Adaptive Interaction}\label{adpative}
To understand the advantages of adaptive integration in OS-Kairos, we compare its performance with and without adaptive integration: when treated as regression optimization or classification optimization in complex scenarios datasets.  As shown in Table~\ref{tab_adpative_interaction},  we see that OS-Kairos without adaptive interaction is quite accurate at HSR of 86.99\%, but its overall performance and intervention precision are suboptimal. For example, the TSR is 82.61\%, IP is 70.66\%, and AP is 95.84\%. The results show that the adaptive interaction does not significantly affect the performance of OS-Kairos. In contrast, OS-Kairos has the advantage of adaptive interaction by tuning the threshold, which balances the sensitivity between autonomous and human intervention. 
\begin{table*}[ht]
    \centering
    \footnotesize
    \begin{tabular}{ccccccc}
      \toprule
      \textbf{Models} & \textbf{Type (\%)}$\uparrow$ & \textbf{SR (\%)}$\uparrow$ & \textbf{TSR (\%)}$\uparrow$ & \textbf{HSR (\%)$\uparrow$} & \textbf{IP(\%)}$\uparrow$& \textbf{AP(\%)}$\uparrow$ \\ \midrule
      OS-Kairos & \textbf{99.88} & \textbf{95.90} & \textbf{88.20} & 86.87 & \textbf{70.75} & \textbf{96.44} \\
      OS-Kairos$_{\text{w/o adaptive interaction}}$ & 99.53 & 95.31 & 82.61 & \textbf{86.99} & 70.66 & 95.84 \\
      \bottomrule
    \end{tabular}
    \caption{Ablation of adaptive interaction.}
    \label{tab_adpative_interaction}
\end{table*}

\section{Case study}
\begin{figure*}[ht]
    \centering
    \includegraphics[width=1\linewidth]{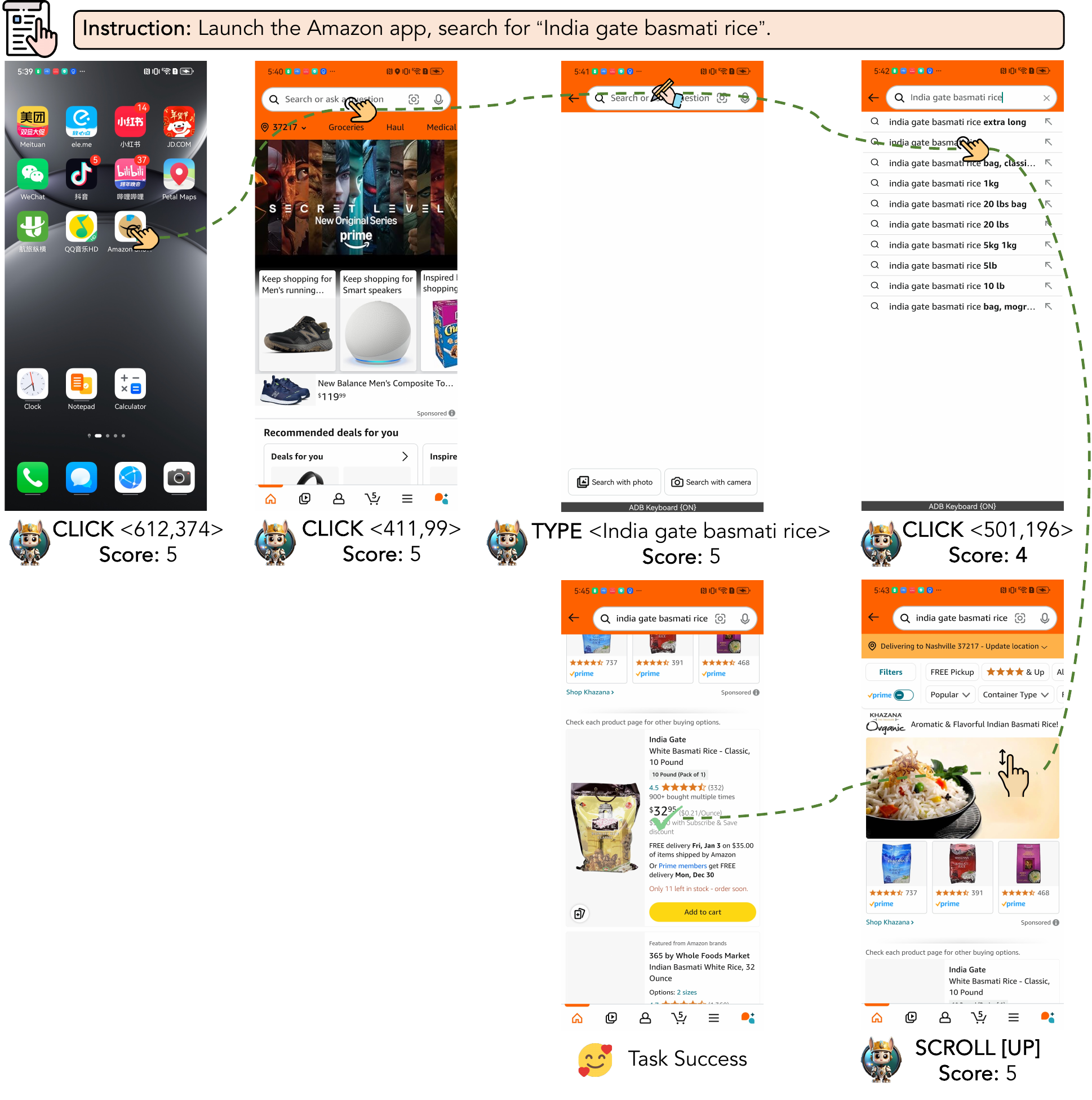}
    \caption{Case study of OS-Kairos in the normal scenario. At each step, OS-Kairos outputs both the action and the confidence score. If the score falls below a specified threshold, human intervention is initiated to ensure task success.}
    \label{normal}
\end{figure*}

\begin{figure*}[ht]
    \centering
    \includegraphics[width=1\linewidth]{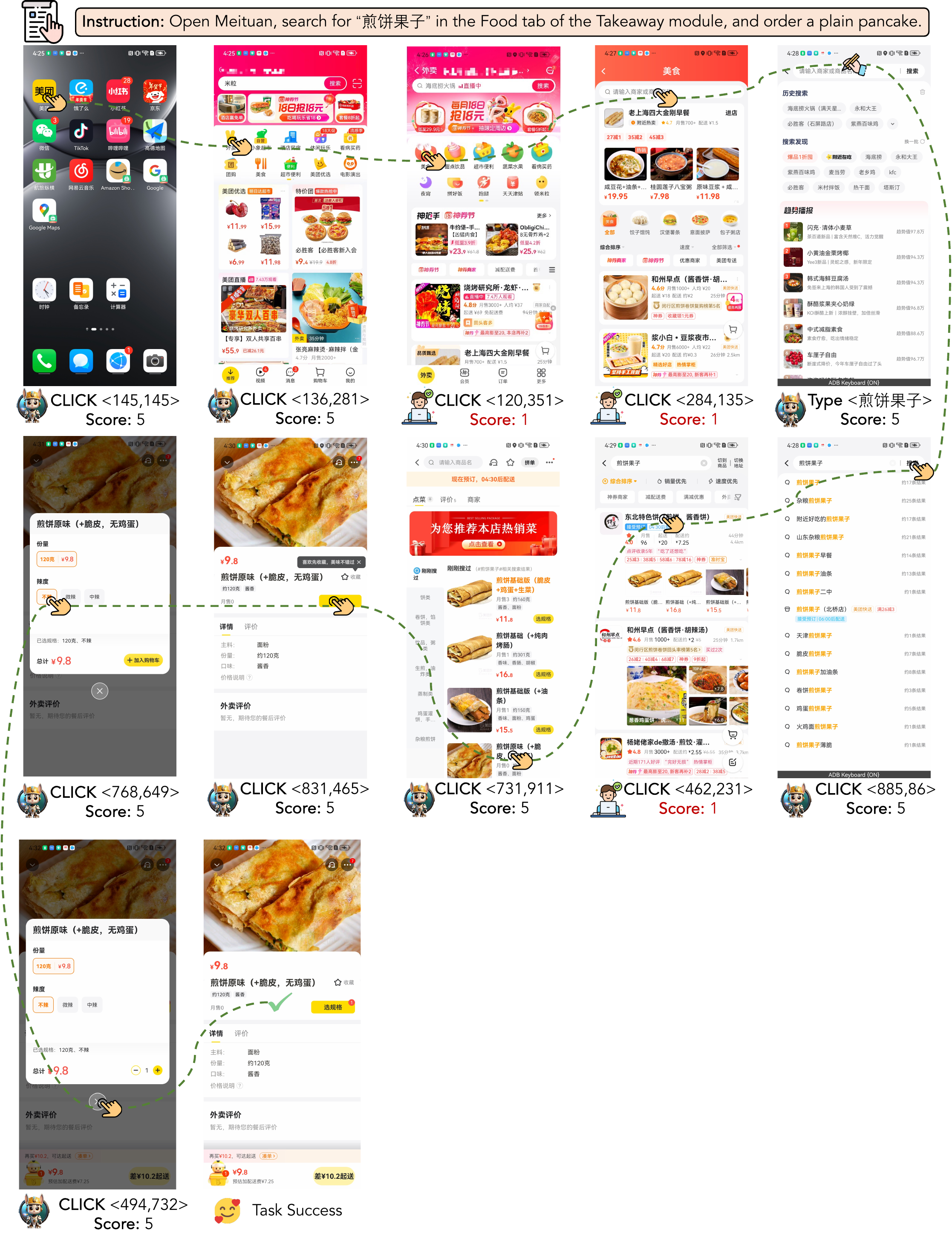}
    \caption{Case study of OS-Kairos in Scenario 1 (capability bottleneck). At each step, OS-Kairos outputs both the action and the confidence score. If the score falls below a specified threshold, human intervention is initiated to ensure task success.}
    \label{scenario1}
\end{figure*}

\begin{figure*}[ht]
    \centering 
    \includegraphics[width=1\linewidth]{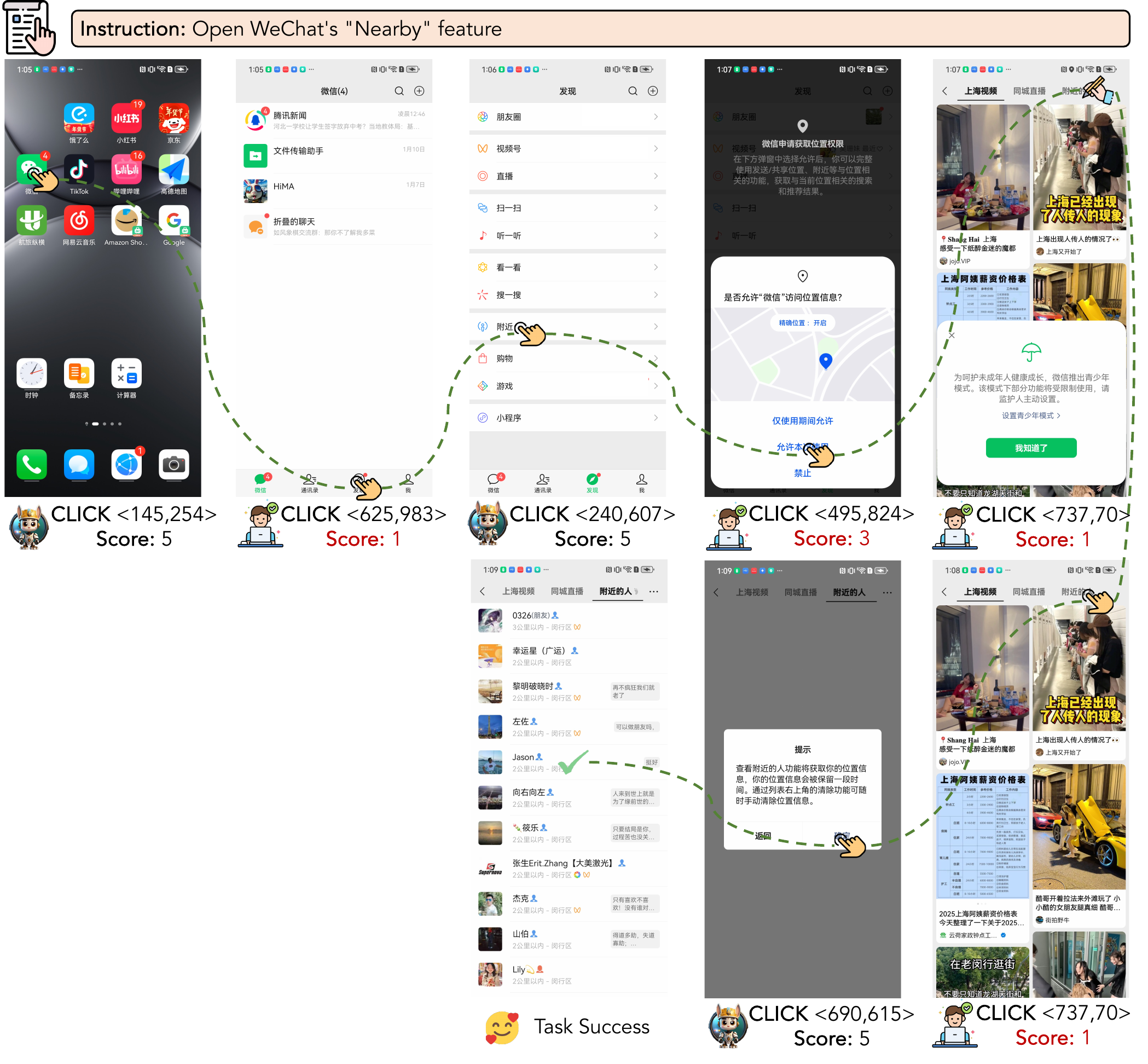}
    \caption{Case study of OS-Kairos in Scenario 2 (no location permission). At each step, OS-Kairos outputs both the action and the confidence score. If the score falls below a specified threshold, human intervention is initiated to ensure task success.}
    \label{scenario2}
\end{figure*}

\begin{figure*}[ht]
    \centering 
    \includegraphics[width=1\linewidth]{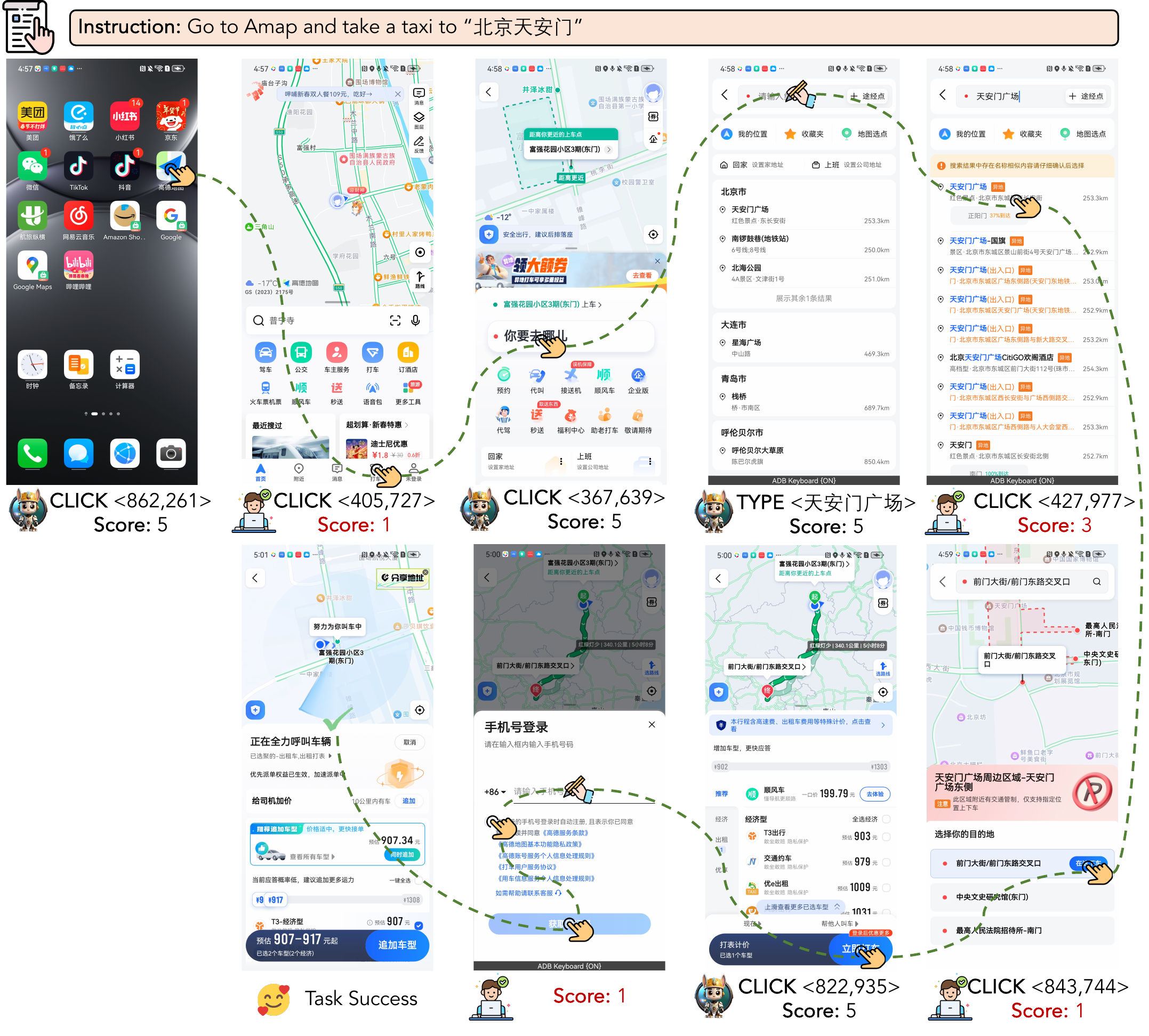}
    \caption{Case study of OS-Kairos in Scenario 3 (information absence). At each step, OS-Kairos outputs both the action and the confidence score. If the score falls below a specified threshold, human intervention is initiated to ensure task success.}
    \label{scenario3}
\end{figure*}

\begin{figure*}[ht]
    \centering 
    \includegraphics[width=1\linewidth]{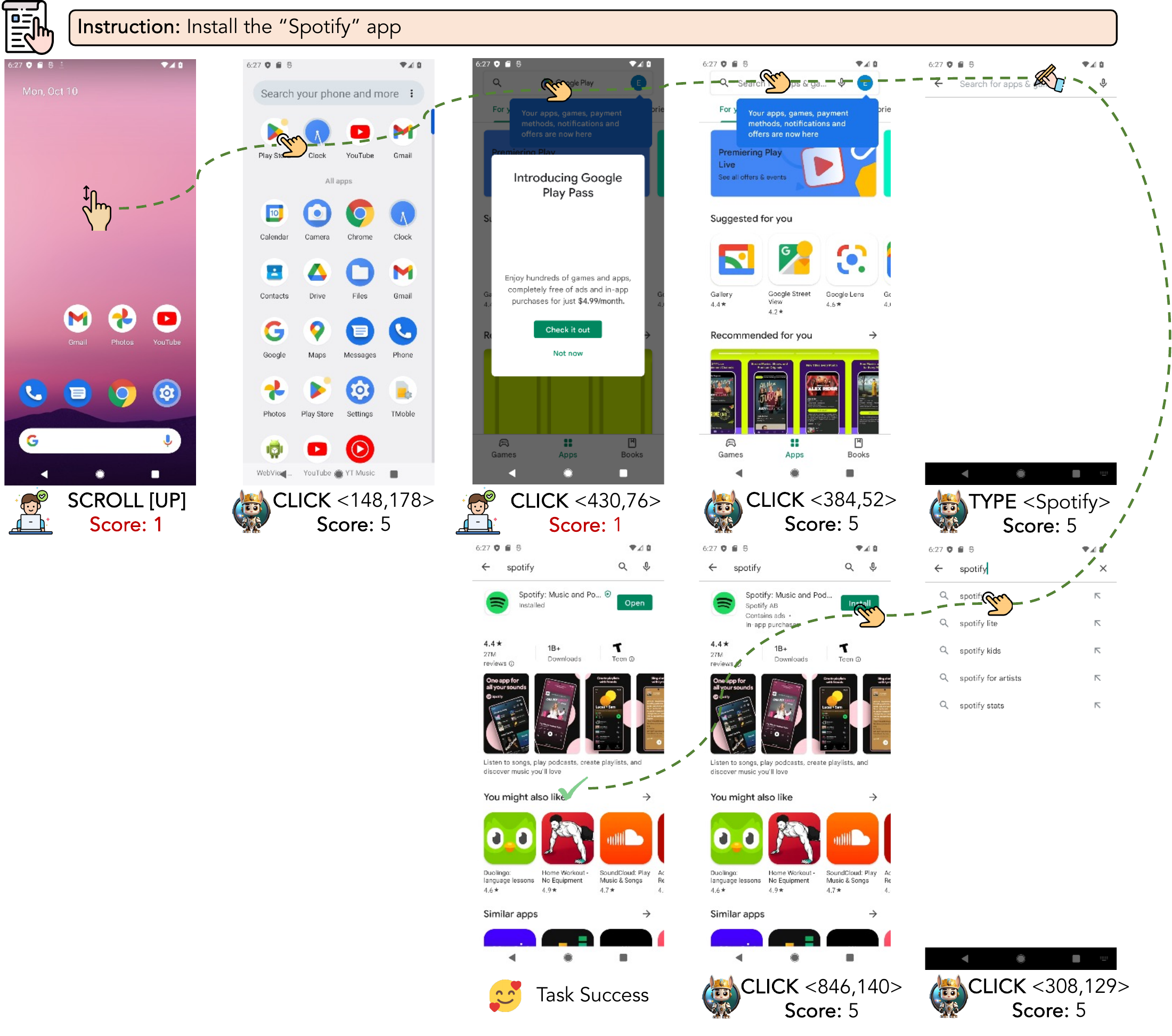}
    \caption{Case study of OS-Kairos in the AITZ benchmark. At each step, OS-Kairos outputs both the action and the confidence score. If the score falls below a specified threshold, human intervention is triggered to ensure task success.}
    \label{aitz_demo}
\end{figure*}

\begin{figure*}[ht]
    \centering 
    \includegraphics[width=1\linewidth]{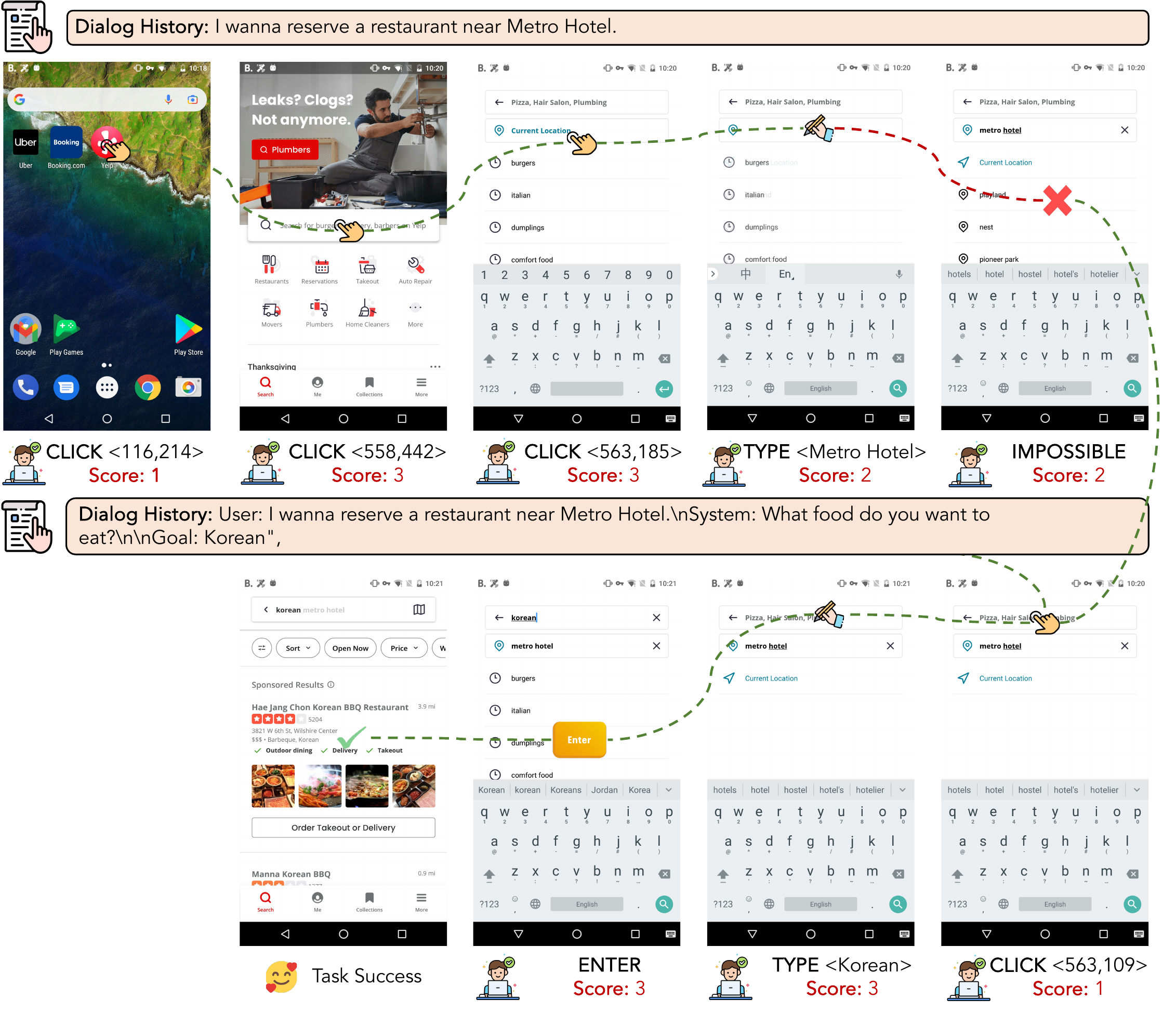}
    \caption{Case study of OS-Kairos in the Meta-GUI benchmark. At each step, OS-Kairos outputs both the action and the confidence score. If the score falls below a specified threshold, human intervention is triggered to ensure task success.}
    \label{meta}
\end{figure*}

To further illustrate the execution process of OS-Kairos, we present four examples from three complex scenarios, along with two examples from the benchmark datasets. First, for simple instructions, OS-Kairos can be fully autonomous, as shown in Figure~\ref{normal}. Second, for complex instructions across the three scenarios, OS-Kairos adaptively identifies the complex steps requiring human intervention, while automating other steps, as shown in Figure~\ref{scenario1}, Figure~\ref{scenario2} and Figure~\ref{scenario3}. Similarly, OS-Kairos performs effectively on the AITZ benchmark (Figure~\ref{aitz_demo}). In an extreme case, OS-Kairos requests human intervention at nearly every step to complete the task, as the Meta-GUI benchmark represents an OOD scenario for OS-Atlas-Pro-7B, as shown in Figure~\ref{meta}.

\end{document}